\PassOptionsToPackage{prologue,table}{xcolor}
\documentclass[sigconf,authorversion]{acmart}
\usepackage{multirow}
\usepackage{subcaption}
\usepackage{cleveref}
\usepackage{array}
\usepackage{longtable}
\usepackage{enumitem}
\usepackage[table]{xcolor}
\usepackage{acmart-taps}

\newlength{\colonewidth}
\newlength{\coltwowidth}
\newlength{\colthreewidth}

\newcommand{\eg}{{e.g.,\ }}

\newcommand{\etc}{{etc.}}
\newcommand{\ie}{{i.e.,\ }}

\def\ReferentoneI{\textit{Signal Awareness}}%
\def\ReferentoneII{\textit{Signal the Robot to Serve Partner}}%
\def\ReferentoneIII{\textit{Do Not Provide Feedback}}%
\def\ReferentoneIV{\textit{Interrupt}}%
\def\ReferenttwoI{\textit{Signal Wrong Drink}}%
\def\ReferenttwoII{\textit{Signal Emergency Stop}}%
\def\ReferenttwoIII{\textit{Indicate Drink Position}}%
\def\ReferenttwoIV{\textit{Signal to Prevent Drink Spilling}}%
\def\ReferenttwoV{\textit{Provide Bad Feedback}}%
\def\ReferentthreeI{\textit{Dismiss}}%
\def\ReferentthreeII{\textit{Call Over}}%
\def\ReferentthreeIII{\textit{Signal the Robot to Collect Cup}}%
\def\ReferentthreeIV{\textit{Provide Good Feedback}}%

\newcommand{\role}{\textsc{ConversationRole}}
\newcommand{\RoleL}{\textit{listener}}
\newcommand{\RoleS}{\textit{speaker}}
\newcommand{\robot}{\textsc{RobotMorphology}}
\newcommand{\AerialTechnical}{\textit{aerial technical}}
\newcommand{\GroundedTechnical}{\textit{grounded technical}}
\newcommand{\Anthropomorphic}{\textit{anthropomorphic}}
\newcommand{\Zoomorphic}{\textit{zoomorphic}}
\newcommand{\referent}{\textsc{Referent}}

\newcommand{\modality}{{Modality}}
\newcommand{\handAndArmGesture}{\textsc{Gesture}}

\newcommand{\verbal}{\textsc{Verbal}}
\newcommand{\eyeGaze}{\textsc{Eye}}
\newcommand{\headMotion}{\textsc{Head}}
\newcommand{\upperBody}{\textsc{UpperBody}}
\newcommand{\legAndFoot}{\textsc{LegAndFoot}}
\newcommand{\facialExpression}{\textsc{Facial}}

\newcommand{\GNI}{\textit{dismissive wave}}
\newcommand{\GNII}{\textit{beckoning wave}}
\newcommand{\GNIII}{\textit{wave}}

\newcommand{\GNIV}{\textit{palm to stop the robot}}

\newcommand{\GNVII}{\textit{thumb up}}
\newcommand{\GNVIII}{\textit{thumb down}}

\newcommand{\GPPII}{\textit{palm point across the table}}
\newcommand{\GPPIII}{\textit{palm point to the place for the drink}}

\newcommand{\GFPII}{\textit{finger point to partner}}

\newcommand{\GNXIII}{\textit{hold the drink}}

\newcommand{\CODEMR}{\textit{repetition}}
\newcommand{\CODEH}{\textit{number-of-hands}}
\newcommand{\HH}{\textit{hand height}}

\newcommand{\VC}{\textit{exact content}}
\newcommand{\VSA}{\textit{speech act}}
\newcommand{\VPronoun}{\textit{unclear reference}}
\newcommand{\VPI}{\textit{Apology}}
\newcommand{\VPII}{\textit{Could You}}
\newcommand{\VPIII}{\textit{First-Person Plural}}
\newcommand{\VPIV}{\textit{Gratitude}}
\newcommand{\VPV}{\textit{Hello}}
\newcommand{\VPVI}{\textit{Please}}
\newcommand{\VPVII}{\textit{Positive Emotion}}
\newcommand{\VPVIII}{\textit{Reasoning}}
\newcommand{\VPIX}{\textit{Subjectivity}}
\newcommand{\VV}{\textit{volume}}
\newcommand{\EGI}{\textit{turn to the robot}}
\newcommand{\EGII}{\textit{turn to the drink}}
\newcommand{\EGIII}{\textit{turn to across the table}}
\newcommand{\EGIV}{\textit{turn to the place for the drink}}
\newcommand{\HGI}{\textit{turn to robot}}
\newcommand{\HGII}{\textit{turn to drink}}
\newcommand{\HGIII}{\textit{turn to across the table}}
\newcommand{\HGIV}{\textit{turn to the place for the drink}}
\newcommand{\HMIV}{\textit{turn to the side}}
\newcommand{\HMI}{\textit{shake}}
\newcommand{\HMII}{\textit{nod}}
\newcommand{\HMIII}{\textit{jaw point to across the table}}
\newcommand{\BI}{\textit{lean away from the robot}}
\newcommand{\BII}{\textit{lean towards the robot}}
\newcommand{\BIII}{\textit{lean back}}
\newcommand{\FEI}{\textit{purse the lips}}
\newcommand{\FEII}{\textit{frown}}
\newcommand{\FEIII}{\textit{raise eyebrow}}
\newcommand{\FEIV}{\textit{smile}}
\newcommand{\LFI}{\textit{stamp}}
\newcommand{\LFII}{\textit{foot draw circle}}
\newcommand{\LFIII}{\textit{foot draw line}}
\newcommand{\LFIV}{\textit{kick away}}

\newcommand{\EGIIW}{\textit{turn to the wrong drink}}
\newcommand{\EGIIC}{\textit{turn to the correct drink}}
\newcommand{\HGIIW}{\textit{turn to wrong drink}}
\newcommand{\HGIIC}{\textit{turn to correct drink}}

\newcommand{\dismissiveWave}{\GNI}
\newcommand{\beckoningWave}{\GNII}
\newcommand{\wave}{\GNIII}
\newcommand{\palmStop}{\GNIV}
\newcommand{\thumbUp}{\GNVII}
\newcommand{\thumbDown}{\GNVIII}
% ...

\newcommand{\repetition}{\textit{repetition}}
\newcommand{\handedness}{\textit{handedness}}
\newcommand{\numberOfHands}{\textit{number-of-hands}}
\newcommand{\handHeight}{\textit{hand height}}

\newcommand{\exactContent}{\VC}
\newcommand{\speechAct}{\VSA}

\newcommand{\volume}{\VV}

\newcommand{\politeness}{{politeness}} % a 'theme', not a code
\newcommand{\politenessApology}{\VPI}
\newcommand{\politenessCouldYou}{\VPII}
\newcommand{\politenessFirstPersonPlural}{\VPIII}
\newcommand{\politenessFormalTitle}{\VPIV}
\newcommand{\politenessGratitude}{\VPV}
\newcommand{\politenessHello}{\VPVI}
\newcommand{\politenessPlease}{\VPVII}
\newcommand{\politenessPositiveEmotion}{\VPVIII}
\newcommand{\politenessReasoning}{\VPIX}
\newcommand{\politenessSubjectivity}{\VPIX}

\definecolor{oxfordblue}{rgb}{0.0, 0.13, 0.28}
\definecolor{harvardcrimson}{rgb}{0.79, 0.0, 0.09}
\definecolor{dartmouthgreen}{rgb}{0.05, 0.5, 0.06}
\definecolor{princetonorange}{rgb}{1.0, 0.56, 0.0}
\definecolor{yaleblue}{rgb}{0.06, 0.3, 0.57}
\definecolor{usccardinal}{rgb}{0.6, 0.0, 0.0}
\definecolor{uclablue}{rgb}{0.33, 0.41, 0.58}
\definecolor{msugreen}{rgb}{0.09, 0.27, 0.23}
\definecolor{cornellred}{rgb}{0.7, 0.11, 0.11}
\definecolor{pomegranate}{RGB}{192, 57, 43}
\definecolor{anti-pomegranate}{RGB}{43,178,192}
\definecolor{alizarin}{RGB}{231, 76, 60}
\definecolor{peter}{RGB}{52, 152, 219}
\definecolor{green}{RGB}{22, 160, 133}
\definecolor{anti-green}{RGB}{160,22,118}
\definecolor{turquoise}{RGB}{26, 188, 156}
\definecolor{pumpkin}{RGB}{211, 84, 0}
\definecolor{anti-pumpkin}{RGB}{0,22,211}
\definecolor{carrot}{RGB}{230, 126, 34}
\definecolor{wisteria}{RGB}{142, 68, 173}
\definecolor{anti-wisteria}{RGB}{99,173,68}
\definecolor{amethyst}{RGB}{155, 89, 182}
\definecolor{sunflower}{RGB}{241, 196, 15}
\definecolor{MAIMIAOLV}{RGB}{85, 187, 138}
\definecolor{charlieblue}{RGB}{34, 66, 148}
\definecolor{darkblue}{RGB}{0, 43, 186}
\definecolor{lightblue}{RGB}{28, 128, 200}
\definecolor{capriblue}{HTML}{1B579C}

\newcommand{\diffdelete}[1]{}
\newcommand{\TODO}[1]{}
\setcopyright{acmlicensed}
\copyrightyear{2025}
\acmYear{2025}
\acmDOI{10.1145/3706598.3714235}
\acmISBN{979-8-4007-1394-1/25/04}
\acmConference[CHI '25]{CHI Conference on Human Factors in Computing Systems}{April 26--May 01, 2025}{Yokohama, Japan}
\acmBooktitle{CHI Conference on Human Factors in Computing Systems (CHI '25), April 26--May 01, 2025, Yokohama, Japan}

\begin{document}
%%
%% The "title" command has an optional parameter,
%% allowing the author to define a "short title" to be used in page headers.
% \title{The Name of the Title Is Hope}
\title[Signaling Human Intentions to Service Robots]{Signaling Human Intentions to Service Robots: Understanding the Use of Social Cues during In-Person Conversations}
% %
% % The "author" command and its associated commands are used to define
% % the authors and their affiliations.
% % Of note is the shared affiliation of the first two authors, and the
% % "authornote" and "authornotemark" commands
% % used to denote shared contribution to the research.

\author{Hanfang Lyu}
\orcid{0000-0003-0135-5754}
\affiliation{\institution{The Hong Kong University of Science and Technology}
\city{Hong Kong}
\country{China}}
\email{hanfang.lyu@connect.ust.hk}

\author{Xiaoyu Wang}
\orcid{0009-0007-6812-7031}
\affiliation{\institution{The Hong Kong University of Science and Technology}
\city{Hong Kong}
\country{China}}
\email{xwangij@connect.ust.hk}

\author{Nandi Zhang}
\orcid{0009-0008-5656-7803}
\affiliation{\institution{University of Calgary}
\city{Calgary}
\state{Alberta}
\country{Canada}}
\email{nandi.zhang@ucalgary.ca}

\author{Shuai Ma}
\orcid{0000-0002-7658-292X}
\affiliation{\institution{The Hong Kong University of Science and Technology}
\city{Hong Kong}
\country{China}}
\email{shuai.ma@connect.ust.hk}

\author{Qian Zhu}
\orcid{0000-0001-5108-3414}
\affiliation{\institution{Renmin University of China}
\city{Beijing}
\country{China}}
\email{zhuqian.vis@gmail.com}

\author{Yuhan Luo}
\orcid{0000-0003-2016-4080}
\affiliation{\institution{City University of Hong Kong}
\city{Hong Kong}
\country{China}}
\email{yuhanluo@cityu.edu.hk}

\author{Fugee Tsung}
\orcid{0000-0002-0575-8254}
\affiliation{\institution{The Hong Kong University of Science and Technology}
\city{Hong Kong}
\country{China}}
\affiliation{\institution{The Hong Kong University of Science and Technology (Guangzhou)}
\city{Guangzhou}
\state{Guangdong}
\country{China}}
\email{season@ust.hk}

\author{Xiaojuan Ma}
\orcid{0000-0002-9847-7784}
\affiliation{\institution{The Hong Kong University of Science and Technology}
\city{Hong Kong}
\country{China}}
\email{mxj@cse.ust.hk}

\renewcommand{\shortauthors}{Lyu, et al.}
\begin{abstract}
    As social service robots become commonplace, it is essential for them to effectively interpret human signals, such as verbal, gesture, and eye gaze, when people need to focus on their primary tasks to minimize interruptions and distractions. Toward such a socially acceptable Human-Robot Interaction, we conducted a study ($N=24$) in an AR-simulated context of a coffee chat. Participants elicited social cues to signal intentions to an anthropomorphic, zoomorphic, grounded technical, or aerial technical robot waiter when they were speakers or listeners. Our findings reveal common patterns of social cues over intentions, the effects of robot morphology on social cue position and conversational role on social cue complexity, and users' rationale in choosing social cues. We offer insights into understanding social cues concerning perceptions of robots, cognitive load, and social context. Additionally, we discuss design considerations on approaching, social cue recognition, and response strategies for future service robots.\enlargethispage{12pt}
\end{abstract}

% %%
% %% Keywords. The author(s) should pick words that accurately describe
% %% the work being presented. Separate the keywords with commas.

\maketitle %%
%% This command processes the author and affiliation and title
%% information and builds the first part of the formatted document.
% \Q{CCS concepts, Keywords}

% \TODO{Check dangling words and sentences, latex packages and warnings, etc.}
% \TODO{Shorten the paper to within 12000 words.}

\section{Introduction}\enlargethispage{12pt}
Service robots have been widely used in public spaces such as retail, healthcare, and hospitality \cite{jonesHumanRobotInteractionUsable2011,ifr2021service}.
Within the human-robot interaction (HRI) community, there have been extensive discussions on how a social robot, as an autonomous and intelligent agent, should behave socially (e.g., \cite{tjomsland2022mind, mcquillinLearningSociallyAppropriate2022}). At the same time, it is equally important to understand how humans intuitively interact with and signal their intentions to robots in social settings \cite{eysselExperimentalPsychologicalPerspective2017}.
Humans have many means to convey their intentions and apply them dynamically according to context. 
For example, when meeting potential employers over coffee, a person may want to ask a drink-serving robot to move aside to avoid potential interruption. Since they need to focus on the ongoing conversation, they might not have enough bandwidth to open the mobile app and send a command or feel embarrassed to speak aloud \cite{mannemExploringCostInterruptions2023}. 
They may prefer simply gesturing to the robot in this situation. 
In other words, when conventional graphical user interfaces (GUI) and conversational interfaces (CUI) are not efficient, service robots should still be able to understand humans' intentions exhibited in other manners. 

In human-human interactions (HHI), people often use different social cues to signal their intentions and emotions to minimize distractions and interruptions to the main tasks. These social cues come in different modalities, such as gaze, gesture, facial expression, body language, and vocal signals \cite{ambadyHistologySocialBehavior2000,kampeHeyJohnSignals2003}.
For example, people often nod to signal awareness of a friend passing by and wave hands to turn down a floor cleaning service during a conversation.
The same interaction preference also occurs in human-robot communications: users are likely to use natural and intuitive social cues to signal their intentions to robots, 
because this is part of human social communication \cite{vinciarelliSocialSignalsTheir2008} and individuals are inclined to treat robots as social actors \cite{nassComputersAreSocial1994}. 
Previous HHI research suggested that the choice of social cues can be affected by people's perceived relationship with the other parties at the scene and their occupancy by the main task \cite{ambadyHistologySocialBehavior2000}. 
Similarly, when interacting with service robots, humans' choices of social signals may be affected by the robot's morphology, social expectations, and norms \cite{fongSurveySociallyInteractive2003}.
% \enlargethispage{12pt}

Backed by the development of voice recognition and vision-based nonverbal cue detection algorithms, many HRI studies explored how humans interact with robots with social cues such as speech \cite{cauchardDroneMeExploration2015}, gesture \cite{cauchardDroneMeExploration2015, firestoneLearningUsersElicitation2019, canutoIntuitivenessLevelFrustrationBased2022}, gaze \cite{liImplicitIntentionCommunication2017}, posture \cite{mccollClassifyingPersonDegree2017}, and movement of different body parts \cite{muehlhausNeedThirdArm2023}, to name a few.
However, most existing works focused on primary interactions between a single user and a robot in lab settings rather than in uninterruptible social scenarios where human-robot interaction is a side task. 
In addition, some of them only analyzed a selected set of social signal modalities or examined different social cues with a specific robot form (e.g., wearable robot arms \cite{muehlhausNeedThirdArm2023} or drones \cite{cauchardDroneMeExploration2015, firestoneLearningUsersElicitation2019}). 
Taken together, there is still a lack of a comprehensive understanding of humans' choice among diverse social cues to express their intentions to different forms of robots in social situations where they are engaging with other people as communicators or respondents. 

In this paper, we explore how humans choose and combine different modalities of social cues to communicate with a service robot during an important social encounter. Specifically, we investigate how robot morphologies and roles in the primary activity influence the use of social cues to express intentions to the robot.
To achieve this, we conducted an elicitation study where participants interacted with a robot server during a simulated coffee chat with potential employers (played by two actors). 
Participants were free to use any social cues they deemed intuitive and appropriate to convey $13$ designated intentions (referents) representing common human feedback types during social interactions with robots.
We selected four representative robots with distinct forms, based on the morphology taxonomy from \cite{onnaschTaxonomyStructureAnalyze2021}: anthropomorphic, zoomorphic, grounded technical, and aerial technical. 
To ensure stability and consistency, we simulated the robots as virtual prototypes \cite{kamideDirectComparisonPsychological2014, muehlhausNeedThirdArm2023} using the augmented reality headset and controlled their actions through a Wizard of Oz (WoZ) approach. 
Each participant ($N=24$) completed two elicitation sessions, alternating between two conversation roles: \RoleS{}, introducing themselves and their experiences, and \RoleL{}, observing the conversation but free to interject. Each session featured a different robot morphology ($2$ out of $4$ per participant), with counterbalanced orders to mitigate learning and order effects.
Following the elicitation sessions, we conducted retrospective think-aloud interviews to understand participants’ mental models and decision-making processes. 
By integrating quantitative analysis of observed social cues with qualitative insights from interviews, we identified patterns in participants’ preferences and rationales.
Participants favored intuitive and context-appropriate cues, such as using eye gaze to signal awareness or hand gestures to guide actions. Robot morphology influenced their choices, particularly in gestures and verbal features, shaped by perceptions of the robot’s sensory capabilities. These findings underscore participants’ goals of minimizing conversational disruptions, ensuring clarity in communication, and maintaining politeness in professional social contexts.

In this paper, we make the following contributions:
\begin{itemize}
    \item We conducted an elicitation study to explore how humans interact with robot waiters on the side during an important coffee chat with potential employers, and how the robot forms and users' conversation roles affect the choice of social cues to signal intentions.
    \item Through retrospective think-aloud and follow-up interviews, we {analyze} the rationales of participants' interactions and {discuss} factors that may influence their intentions when choosing social cues.
    \item Based on our findings, we further {discuss} the design implications for the human-robot interaction system in social settings and {provide} suggestions for future elicitation research for human-robot interactions.
\end{itemize}

\section{Related Work}
\label{sec:related-work}

\subsection{Service Robots in Social Settings}

\subsubsection{Definitions and Scope}
Social settings are systems centered on \textit{social processes} (\ie interactions between two or more individuals) structured by \textit{resources} and the \textit{organization of resources} \cite{tsengSystemsFrameworkUnderstanding2007}.
In this study, we refer to social encounters where human-human interaction dominates, such as a coffee chat, a group discussion, or a dinner party, as social processes,
with social service robots acting as resources to support the social processes.
Social robots, as defined by \citet{yanSurveyPerceptionMethods2014}, are ``robots which can execute designated tasks, and the necessary condition turning a robot into a social robot is the ability to interact with humans by adhering to certain social cues and rules''.
They share key features such as sensing and responding to environmental cues, interacting with humans (or other robots), and understanding and following social rules \cite{sarricaHowManyFacets2019}.
\citet{yanSurveyPerceptionMethods2014} also emphasize the importance of recognition capabilities and social cues for social service robots.

\subsubsection{Challenges in Designing Intuitive HRI in Social Settings}
Human interactions with social robots present unique challenges that distinguish it from general human-robot interaction (HRI). 
The ability of social service robots to understand and respond to users' intentions and preferences has been considered critical to ensure safety, human control, and alignment with human expectations and preferences \cite{sharkawyHumanRobotInteraction2022,stephanidisSevenHCIGrand2019}.
\citet{tianTaxonomySocialErrors2021} present a comprehensive taxonomy of social errors in HRI, illustrating the complexity of social interactions between humans and robots. 
This complexity extends beyond mere functionality to encompass aspects such as timing, appropriateness, and emotional congruence.
Key considerations in social robot interactions include social appropriateness and adaptability \cite{tjomsland2022mind, mcquillinLearningSociallyAppropriate2022}, emotional and affective aspects \cite{kirby2010affective}, temporal dynamics \cite{stedtlerThereReallyEffect2024}, error handling and repair \cite{lee2024rex}, and context-specific behaviors \cite{law2022friendly}.
While prior work has investigated how robots perceive and respond to human behavior and highlighted the importance of properly understanding humans in social HRI, 
there is a limited understanding of how humans intuitively interact with service robots in social encounters. 
In the social settings of our study, robots are positioned in a side task, providing services to the participants rather than central actors among humans. 
We examine how humans naturally communicate with robots in such contexts, 
aiming to better interpret human signals, inform the design of social service robots, and provide intuitive services to human-centric social encounters.
% \enlargethispage{12pt}

\subsection{Social Cues in Human-Robot Interaction}

\subsubsection{Human Social Cues}
Social Signal Processing (SSP) analyzes the social behaviors and cues in both Human-Human (HHI) and Human-Computer interactions (HCI) contexts \cite{vinciarelliSocialSignalsTheir2008}. 
\citet{vinciarelliSocialSignalsTheir2008,vinciarelliBridgingGapSocial2012} provide a comprehensive summary of human nonverbal behavior cues, their functions and social behavior modeling.
Multimodal analysis of nonverbal behaviors in social interactions includes applications in modeling multimodal behaviors for face-to-face social interaction \cite{mihoubLearningMultimodalBehavioral2015}, automatic categorization of autism spectrum disorder \cite{chenComputingMultimodalDyadic2017}, and classifying perceptions of interdependence \cite{dudzikRecognizingPerceivedInterdependence2021}.
The relationship between human social cues and emotions is also widely explored \cite{glowinskiBodySpaceEmotion2017, luoEmotionEmbodiedUnveiling2024}.
However, much of the work in SSP has focused on the analysis and processing of human behaviors, with a limited understanding of how humans select and adapt social cues when interacting with robots in dynamic social settings.
While social cue processing for robots has advanced, focusing on how robots express their own social intentions through gaze \cite{moonMeetMeWhere2014,diethelmEffectsGazeSpeech2021}, speech \cite{diethelmEffectsGazeSpeech2021}, gestures \cite{kwonExpressingRobotIncapability2018,matareseRobotsBehavioralTransparency2021}, and facial expressions \cite{broekensTransparentRobotLearning2021,matareseRobotsBehavioralTransparency2021}, 
less attention has been paid to how humans intuitively communicate their intentions to social service robots, particularly in complex, real-world settings where robots play a peripheral role.
Our study addresses this gap through an exploratory study focusing on the variety of social cues humans employ to convey different intentions in interactions with service robots during social encounters.

\subsubsection{Social Cues Elicitation in Human-Robot Interaction}
Research in HCI has been exploring how humans interact with various systems using different social cues, such as public displays \cite{rodriguezGestureElicitationStudy2017}, smart rings \cite{gheranGesturesSmartRings2018}, and chairs \cite{andreiTakeSeatMake2024}. 
\citet{villarreal-narvaezSystematicReviewGesture2020} systematically reviewed the literature on gesture elicitation studies in HCI.
Furthermore, studies on HRI have also examined the use of different social cues to interact with robots.
\citet{cauchardDroneMeExploration2015} explored how humans interact with drones with gesture and speech, and \citet{firestoneLearningUsersElicitation2019} collected elicitation of gestures for small Unmanned Aerial Systems (sUAS) to understand and model human-drone interactions.
\citet{canutoIntuitivenessLevelFrustrationBased2022} proposed a frustration-based elicitation approach and studied the intuitiveness of human gestures to signal robots with basic commands.
However, most of these studies have focused on primary interactions between a single user and a robot in controlled, lab-based settings and have neglected more complex, uninterrupted social scenarios in which HRI occurs as a side task.
Furthermore, these studies limit their focus to a narrow range of social signal modalities or specific robot forms (\eg{} drones \cite{cauchardDroneMeExploration2015, firestoneLearningUsersElicitation2019} or wearable robot arms \cite{muehlhausNeedThirdArm2023}), failing to provide a comprehensive understanding of the variety of social cues employed in dynamic, real-world HRI scenarios.
We aim to explore human interactions with multiple forms of service robots when occupied by primary social encounters, to elicit social cues that cover all possible modalities from human bodies, and to understand the choice of social cues to interact with service robots as a peripheral task, expanding the scope of social cue elicitation research in HRI.
% \enlargethispage{12pt}
\subsection{Morphology of Social Service Robots}

\subsubsection{Classification of Robot Morphology}
Robot morphology is one of the fundamental classification parameters in HRI research \cite{fongSurveySociallyInteractive2003,yancoClassifyingHumanrobotInteraction2004,onnaschTaxonomyStructureAnalyze2021}.
As refined by \citet{onnaschTaxonomyStructureAnalyze2021}, robot morphology can be classified as \textit{anthropomorphic} (human-like and androids), \textit{zoomorphic} (animal-like), or \textit{technical} (machine-like).
Robot morphology determines a robot's physical embodiment and influences users' perceptions of its functional and communicative capabilities \cite{onnaschTaxonomyStructureAnalyze2021}.
\citet{eysselExperimentalPsychologicalPerspective2017} provides an experimental psychological perspective on social robotics, emphasizing how human cognitive and social psychological processes influence perceptions of and interactions with social robots.
In our experiment, we apply \citet{onnaschTaxonomyStructureAnalyze2021}'s taxonomy of robot morphology, aiming to investigate how the visual design of a robot influences human perceptions of its embodiment and capabilities and human usages of social cues during interactions with service robots in social settings.

\subsubsection{Morphologies of Social Service Robots}

Social robots have been widely applied in service industries \cite{jonesHumanRobotInteractionUsable2011,ifr2021service,mcquillinLearningSociallyAppropriate2022} with various morphologies, including \textit{anthropomorphic} \cite{leeRippleEffectsEmbedded2012,curtisPizzaHutHires2016,RobotsServeFood2019}, \textit{zoomorphic} \cite{TouristsAmazedChina,jacksonRobotDogWaiter2020,dossantosZooVisitorsInitial2023} and \textit{technical} \cite{WaiterDroneVideo,SingaporeRestaurantShows2015,RobotWaiterRestaurant,PuduBot2PuduRobotics} robots.
The \textit{technical} morphology of service robots, \ie product-oriented robots, is most commonly seen among the industrial applications of service robots, such as drones \cite{WaiterDroneVideo,SingaporeRestaurantShows2015,kongEffectsHumanConnection2018}, cleaning robots \cite{RoombaRobotVacuum}, and delivery robots \cite{RobotWaiterRestaurant,wanWaiterRobotsConveying2020}.
Compared with the \textit{technical} morphology, research has been exploring how \textit{anthropomorphic} robots affect users' perceptions and interactions with social robots.
\citet{kwakImpactRobotAppearance2014} compare the social presence and sociability of a human-oriented robot and a product-oriented robot.
\citet{stroessnerSocialPerceptionHumanoid2019} examine the effects of gendered and machine-like features on the social perception of humanoid and non-humanoid robots.
Most research of \textit{zoomorphic} social service robots has been focused more on companion and guidance applications \cite{marchettiPetRobotApplianceCare2022,dossantosZooVisitorsInitial2023,hwangRoboticCompanionsUnderstanding2024}.
\citet{hauserWhatsThatRobot2023} explore human perceptions of incidental encounters with service robot dogs in a lab setting, suggesting a positive experience for the human.
Existing research on three morphologies of social service robots has been limited to academic settings, with few empirical results on the effects of robot morphology on human interactions with robots, especially in real-world social settings.
Thus, we add robot morphology as an independent variable in our study to explore the effects of robot morphology on human interactions with social service robots.

\subsection{Exploratory Prototyping in Human-Robot Interaction}
\label{sec:related-work-virtual-prototyping}

Exploratory prototyping plays a crucial role in HRI research, enabling researchers to rapidly test and assess the feasibility and usability of experimental robot designs \cite{zamfirescu-pereiraFakeItMake2021}.
Research by \citet{rojasComparisonRobotAssessment2024} indicates that virtual and physical robot prototypes are comparable in observable aspects (e.g., color and shape) and emotional responses, as measured by the Self-Assessment Manikin instrument, but differ in social perceptions, such as discomfort and warmth.
Beyond image and video prototyping, virtual reality (VR) has been adopted as a simulation tool for social robots, offering immersive testing environments \cite{shariatiVirtualRealitySocial2018,sadkaVirtualrealitySimulationTool2020}.
For instance, \citet{kamideDirectComparisonPsychological2014} indicates that while VR robots and physical robots may elicit differing subjective impressions, participants' behavior and desired personal space remain consistent between the two prototyping approaches.
Additionally, studies have demonstrated the feasibility of using augmented reality (AR) to collect human social cues in real-world interaction contexts \cite{panWhyHowUse2018,mahadevanGripthatthereInvestigationExplicit2021,huPose2GazeEyeBodyCoordination2024}.
Our primary objective is to gather participants' social cues—specifically, their behaviors when interacting with service robots. 
Given that virtual robot prototypes effectively replicate key aspects of interaction observed with physical robots \cite{kamideDirectComparisonPsychological2014}, the use of augmented reality for prototyping is a practical and effective choice for our study.
% \enlargethispage{12pt}

\section{Method}

This section describes our mixed-methods study, including a pilot and elicitation study within a simulated public coffee-chat scenario featuring four service robot morphologies and contextual referents. 
Behavioral and interview data were processed through social cue coding, statistical methods, and thematic analysis of interviews.
All experiments received approval from the Institutional Review Board (IRB) approval from the University Research Ethics Committee.

\subsection{Simulation of Robots and Public Context Settings}
\subsubsection{Social Encounter}
\label{method: simulation: social-encounter}
Our study examines human interaction with the robot as a side task, positioning the social encounter, \ie human-human interaction, as the primary task to be interleaved by robots in our social settings.
Conversation has been a central topic in the analysis of human-human interactions \cite{goodwinConversationAnalysis1990,vinciarelliOpenChallengesModelling2015,porcheronUsingMobilePhones2016}.
Thus, we chose conversation as the main task in our experiment settings.
To select an appropriate social scenario for our experiment, we have the following requirements: First, it should involve the participation of multiple people. Second, interruption is unwanted and should be minimized in such social scenarios, but all communication channels should not be fully occupied, allowing users to maintain sufficient options when utilizing social signals.
Based on the above considerations, five researchers brainstormed appropriate scenarios for our experiment and selected a coffee chat with potential employers.
Considering that users' choice of social cues when interrupted by a robot may vary based on their ongoing activity, particularly their conversational role as listener or speaker, we included these roles as an independent variable (IV) \role{} in our experimental design.
The \role{} is simulated by letting the participants hold the conversational flow (\RoleS{}), or mainly listen to the conversation led by the potential employers (\RoleL{}).
We recruited actors to play the roles of two potential employers (advisors) for future jobs, internships, or advanced study according to the participant's status and goals.
The actors were trained to pay attention to the participants and create more chances for the participants to continue their speech when participants were in the \RoleS{} session, and to take the dominance of the conversation when participants were in the \RoleL{} session.
Natural turns of conversation and attention were maintained during the coffee chat to ensure the authenticity of the social encounter.
\begin{table*}[htbp]
    \centering
    \caption{Five Necessary Interaction Types for HRI in Social Settings}
    \label{tab:taxonomy-of-interaction}
    \begin{tabular}{ccl}
        \toprule
        \multirow{2.3}{*}{Robot Active Seeking for Human Input}
         &
        \multicolumn{2}{c}{i. When the robot is not sure}
        \\
        \cmidrule(l){2-3}
         &
        \multicolumn{2}{c}{ii. When the robot asks for evaluation}
        \\
        \midrule
        \multirow{3.6}{*}{Robot Passive Receiving Human Input}
         &
        \multicolumn{2}{c}{iii. When the human signals awareness}
        \\ \cmidrule(l){2-3}
         &
        \multicolumn{1}{c}{\multirow{2.3}{*}{When the robot has an error \cite{tianTaxonomySocialErrors2021}}}
         &
        iv. Performance error
        \\ \cmidrule(l){3-3}
         &
        \multicolumn{1}{c}{}
         &
        v. Social error
        \\
        \bottomrule
    \end{tabular}
\end{table*}
\begin{table*}[htbp]
\centering
\caption{13 Referents and Their Types Used in Our Experiment}
\label{tab:referents-used-in-our-experiment}
\begin{tabular}{lll}
\toprule
\textbf{Referents} &\textbf{Situation} & \textbf{Type} \\ \midrule
\ReferentoneI&The robot was moving to you. & iii \\ \cmidrule{1-3}
\ReferentoneII&The robot stopped near you, carrying a bottle of drink you had just ordered. & v \\ \cmidrule{1-3}
\ReferentoneIII&The robot sent the drink to your partner, and asked ``Please rate my service.'' & ii \\ \cmidrule{1-3}
\ReferentoneIV&The robot was saying ``Ok, your next cup of drink is expected to arrive at ...'' & v \\ \cmidrule{1-3}
\ReferenttwoI&The robot went away then came back, but brought a wrong drink to you. & iv \\ \cmidrule{1-3}
\ReferenttwoII&The robot encountered a malfunction and was rushing towards you. & iv \\ \cmidrule{1-3}
\ReferenttwoIII&The robot went near you and asked ``Where should I place the drink?'' & i \\ \cmidrule{1-3}
\ReferenttwoIV&The robot was sending out the drink, but the drink was spilling. & iv \\ \cmidrule{1-3}
\ReferenttwoV&The robot finished sending the drink and asked ``Please rate my service.'' & ii \\ \cmidrule{1-3}
\ReferentthreeI&The robot was wandering around nearby, disturbing your conversation. & v \\ \cmidrule{1-3}
\ReferentthreeII&Your partner just finished her drink, and wanted the robot to collect the cup. & iii \\ \cmidrule{1-3}
\ReferentthreeIII&The robot moved to you and asked ``How may I help you?'' & i \\ \cmidrule{1-3}
\ReferentthreeIV&The robot collected your partner's cup, and asked ``Please rate my service.'' & ii \\ \bottomrule
\end{tabular}
\end{table*}

\subsubsection{Robot}
\label{method:simulation:robot}
\begin{figure}[htbp]
    \centering
    \begin{subfigure}{.23\textwidth}
        \centering
        \includegraphics[width=\textwidth]{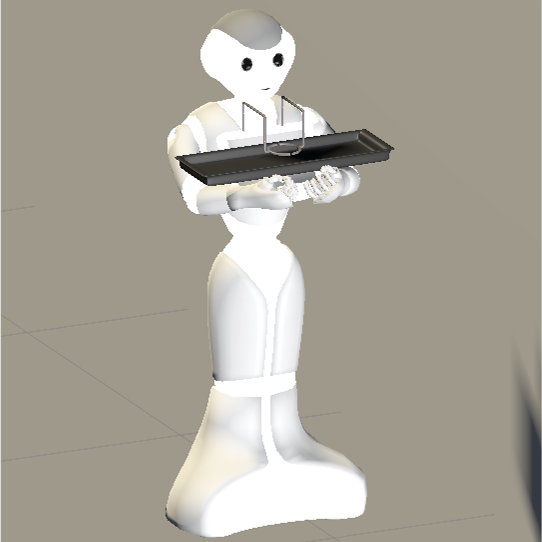}
        \subcaption{Anthropomorphic Robot}
        \label{fig:robot-form-screenshot-humanoid}
    \end{subfigure}
    \hfill
    \begin{subfigure}{.23\textwidth}
        \centering
        \includegraphics[width=\textwidth]{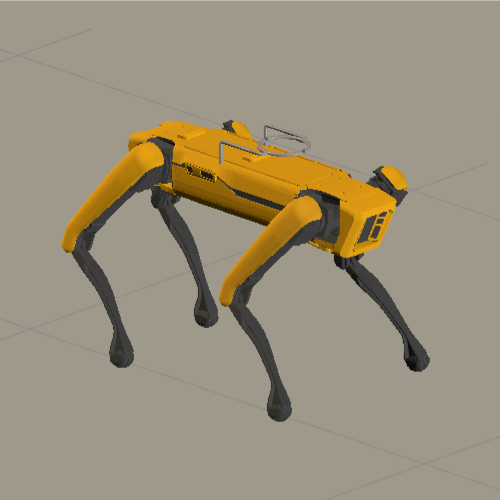}
        \subcaption{Zoomorphic Robot}
        \label{fig:robot-form-screenshot-dog}
    \end{subfigure}
    \hfill
    \begin{subfigure}{.23\textwidth}
        \centering
        \includegraphics[width=\textwidth]{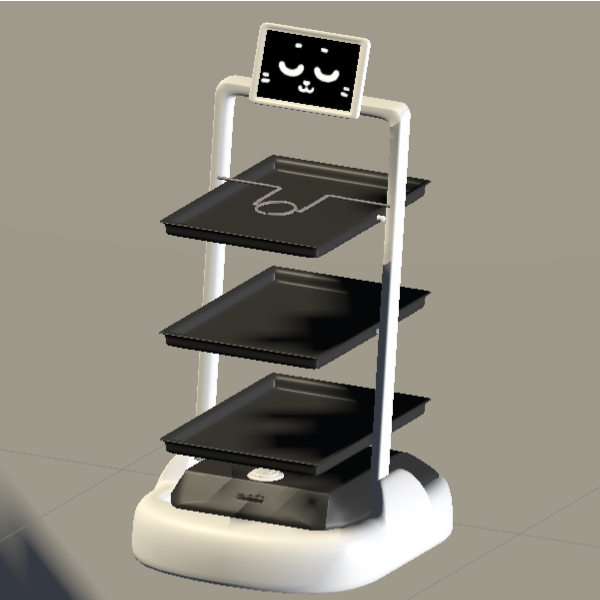}
        \subcaption{Grounded Technical Robot}
        \label{fig:robot-form-screenshot-waiter}
    \end{subfigure}
    \hfill
    \begin{subfigure}{.23\textwidth}
        \centering
        \includegraphics[width=\textwidth]{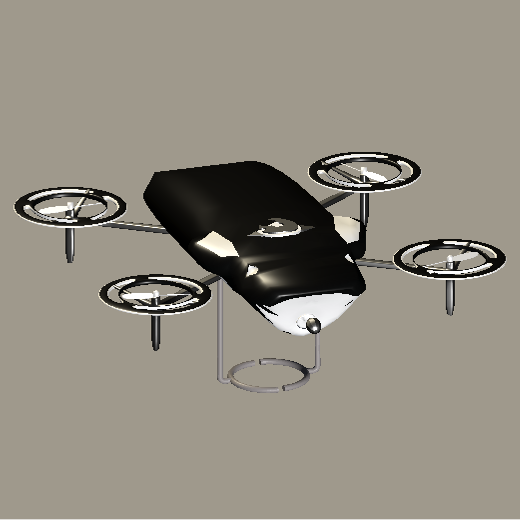}
        \subcaption{Aerial Technical Robot}
        \label{fig:robot-form-screenshot-drone}
    \end{subfigure}
    \caption{Four Forms of Robots Used in Our Experiment}
    \label{fig:robot-form-screenshot}
    \Description{The Four Forms of Robots Used in Our Experiment: (a) Anthropomorphic Robot; (b) Zoomorphic Robot; (c) Grounded Technical Robot; (d) Aerial Technical Robot.}
\end{figure}
We also set the robot morphology as our independent variable (IV), denoted as \robot{}. 
To select the robot morphologies, we initially adopted the taxonomy proposed by \cite{onnaschTaxonomyStructureAnalyze2021}, which classifies robot morphologies into three categories: \textit{technical}, \textit{anthropomorphic}, and \textit{zoomorphic}.
Given that the height of a robot may influence user interaction \cite{hiroiInfluenceHeightRobot2016,raeInfluenceHeightRobotmediated2013} and perception \cite{jostExaminingEffectsHeight2019}, and considering the prevalence of drones
% \enlargethispage{12pt}
\cite{bloombergoriginalsDronesWaitersSushi2013,SingaporeRestaurantShows2015,WaiterDroneVideo} and food delivery carts \cite{BellaBotPuduRobotics,PuduBot2PuduRobotics,RobotWaiterRestaurant} in catering services,
we further divided technical robots into two subcategories: aerial robots (drones) and grounded robots (food delivery carts).
Thus, the IV \robot{} had four experiment groups: \Anthropomorphic{}, \Zoomorphic{}, \GroundedTechnical{}, and \AerialTechnical{} robots.
Different morphologies of robots were simulated using Augmented Reality (AR) technology to provide participants with an immersive and interactive experience while maintaining flexibility in testing different robot morphologies.
As discussed in \Cref{sec:related-work-virtual-prototyping}, humans behave similarly toward virtual and real robots, making AR simulation a suitable choice for prototype experimental settings.
This approach provides valuable insights into understanding users' behavior and informing robot design.
Virtual simulation also ensures consistency across conditions, reduces logistical complexity, and allows for convenient human social cue collection.
To identify representative robots, we first compiled a diverse set of examples for each morphology.
Five researchers independently voted for up to three robots per category based on the following criteria: (1) popularity, (2) ability to serve drinks, (3) perceived safety for novice users, and (4) representativeness of the morphology. 
Based on the voting results, we selected Pepper\footnote{\url{https://us.softbankrobotics.com/pepper}} (Figure~\ref{fig:robot-form-screenshot-humanoid}) as the anthropomorphic representative and Spot\footnote{\url{https://bostondynamics.com/products/spot/}} (Figure~\ref{fig:robot-form-screenshot-dog}) as the zoomorphic representative. For the technical robots, we created a grounded food delivery cart mesh commonly seen in coffee shops (Figure~\ref{fig:robot-form-screenshot-waiter}) and selected a drone equipped with safety measures from the Unity asset store\footnote{\url{https://assetstore.unity.com/packages/3d/vehicles/air/simple-drone-190684}} (Figure~\ref{fig:robot-form-screenshot-drone}).\enlargethispage{12pt}

\subsubsection{Referents}
\label{method:simulation:referents}
We aimed to design effective elicitation referents for human-robot interactions in social encounters by first categorizing the types of necessary interactions between humans and robots.
A literature survey was conducted to summarize the situations where robots require human interaction \cite{stephanidisSevenHCIGrand2019,mehtaUnifiedLearningDemonstrations2024}, providing a structured foundation for the elicitation referents.
Based on this review, five necessary interaction types were identified, as shown in \Cref{tab:taxonomy-of-interaction}. % below.
In our coffee chat scenario, we brainstormed and carefully designed $13$ elicitation referents, covering each of the identified interaction types. 
These referents were designed to reflect common interactions between human waiters and customers and to address the important tasks of a robot waiter as outlined in \cite{garcia-haroServiceRobotsCatering2021}. 
\Cref{tab:referents-used-in-our-experiment} details the elicitation referents, their associated interaction situations, and their classification by interaction type.
To create a realistic social encounter, we situated our study in a shared public workspace with low-volume background music and ambient noise, conditions typical of coffee chats.
This setting mirrors natural environments where human-human social encounters and interactions with service robots frequently occur, ensuring ecological validity.
We programmed the robots to respond to each interaction situation with appropriate actions and sounds in the AR headset for participants.
The robots' actions were controlled using a Wizard of Oz (WoZ) approach, enabling a human operator (wizard) sitting at another table behind the participants to manage the robots' responses in real-time, thereby simulating seamless and context-appropriate interactions.
And since our actors were familiar with the robots' routines and actions, they could infer the progress of the virtual robot and act as if they could see the robots during the referent elicitation, which helped to maintain the authenticity of the AR simulation of the scenario. \enlargethispage{12pt}

\begin{table*}[htbp]
    \centering
    \caption{Participants' familiarities with the four robot morphologies used in our experiment.}
    \label{tab:participant-familiarity}
    \begin{tabular}{ccccc}
        \toprule
         & \AerialTechnical
         & \Anthropomorphic
         & \GroundedTechnical
         & \Zoomorphic
        \\ \midrule
        \textbf{Never heard of}
         & $0 ~ (0.00\%)$
         & $1 ~ (8.33\%)$
         & $0 ~ (0.00\%)$
         & $3 ~ (25.00\%)$
        \\ \midrule
        \textbf{Heard of but never interacted with}
         & $3 ~ (25.00\%)$
         & $8 ~ (66.67\%)$
         & $4 ~ (33.33\%)$
         & $8 ~ (66.67\%)$
        \\ \midrule
        \textbf{Interacted with}
         & $9 ~ (75.00\%)$
         & $3 ~ (25.00\%)$
         & $8 ~ (66.67\%)$
         & $1 ~ (8.33\%)$
        \\ \bottomrule
    \end{tabular}
\end{table*}

\subsubsection{Pilot Study}
To validate our design, we conducted a pilot study ($N=7$) where participants were asked to use social cues to signal the AR-simulated robot in a coffee chat with potential employers under different combinations of conditions: $4$ robot morphologies (\AerialTechnical{}, \GroundedTechnical{}, \Anthropomorphic{}, \Zoomorphic{}) $\times$ $2$ conversation roles (\RoleS{}, \RoleL{}) $\times$ $2$ postural configurations (\textit{sitting around a table}, \textit{standing around a long bar}). 
Each participant completed four sessions, which included two robot forms, both roles, and both postural configurations. 
For each referent, participants were required to elicit three times and make each elicitation as distinct from the others as possible.
According to the post-experiment interview, we improved our study design in the following aspects: 
First, all participants found the requirement of eliciting three times during conversation to be too cognitively demanding. 
Thus, we reduced the three compulsory social cues to one, and participants were free to provide alternatives either during the elicitation or in the interview. 
Second, when asked about the differences between sitting and standing, all participants believed that the effects were much smaller than those caused by roles, except for the emergency case (referent \ReferenttwoII{}). 
Considering that sitting is more common for a coffee chat scenario and for the participants' comfort, we only kept the sitting setting. %, and the formal study was thus reduced from 4 to 2 sessions. 
Finally, we improved the AR simulation experience according to the participants' feedback, such as more authentic robot sound effects, approaching directions, and interaction distances. \enlargethispage{12pt}

\begin{figure*}[htbp]
    \centering
    \begin{subfigure}[t]{.245\textwidth}
        \centering
        \includegraphics[width=\linewidth]{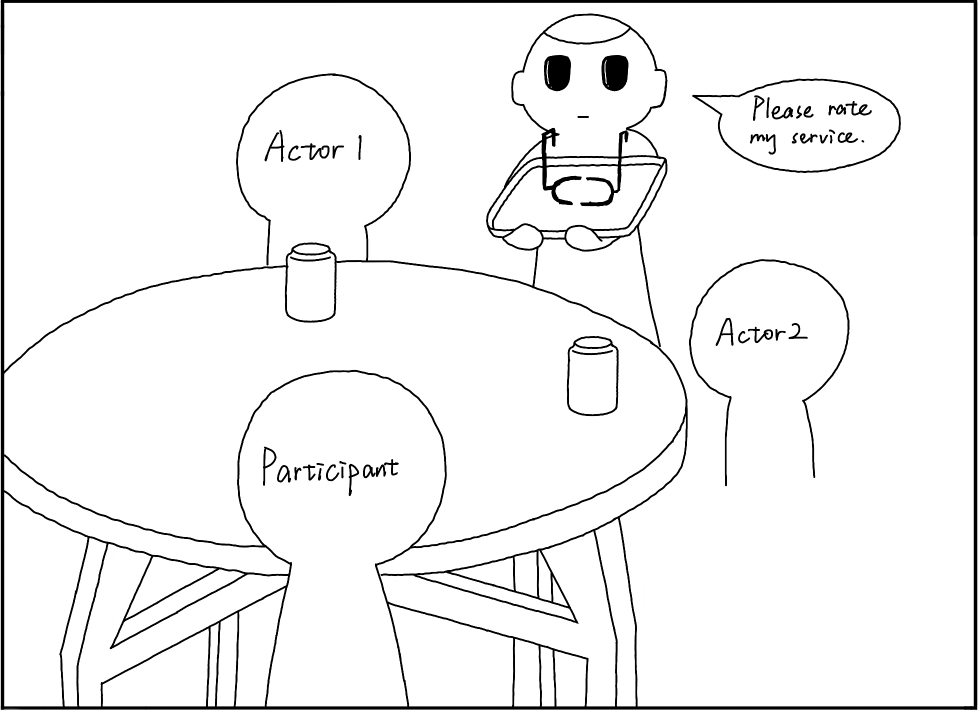}
        \medbreak
        \includegraphics[width=\linewidth]{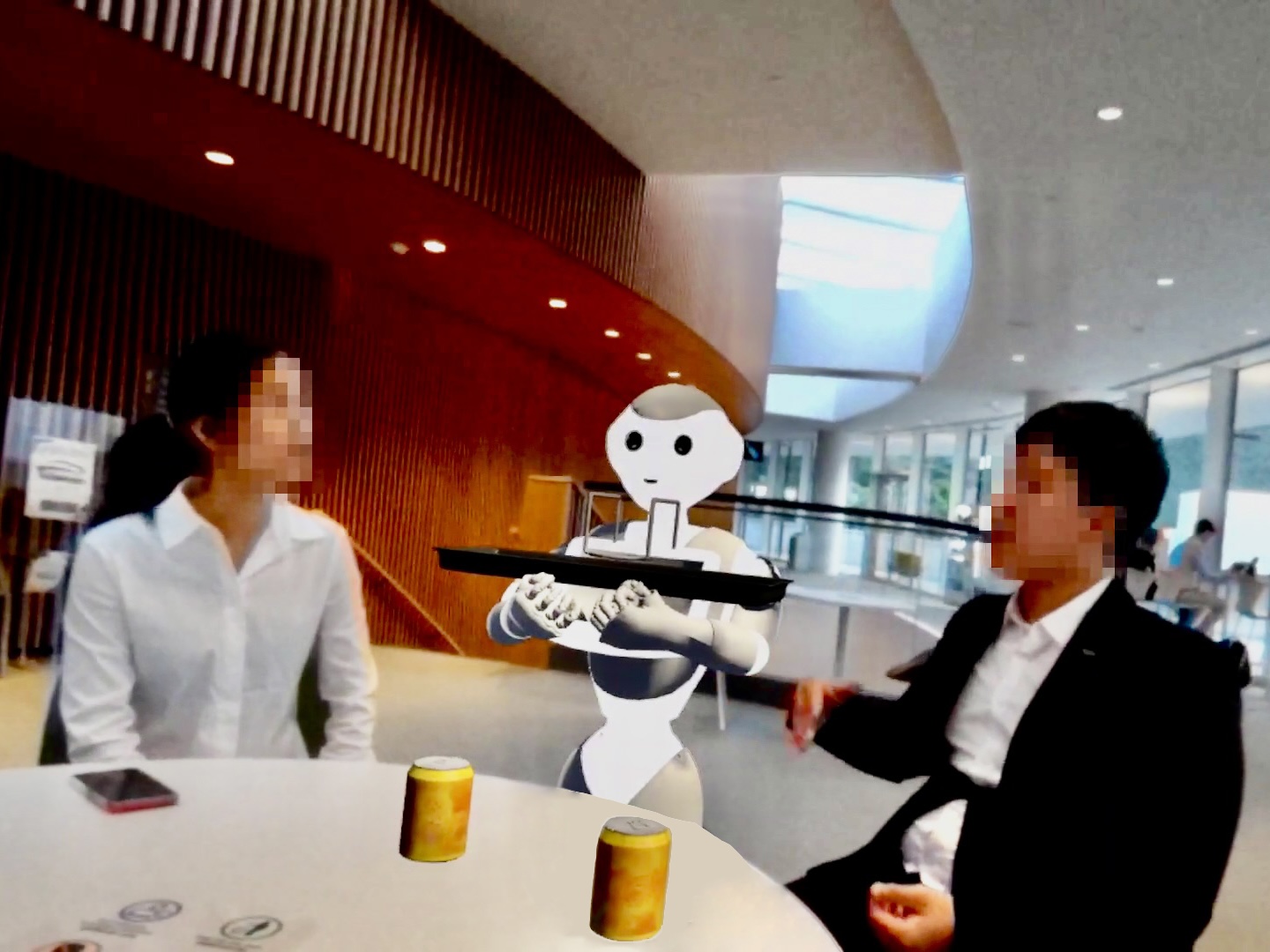}
        \subcaption{Pepper, the \Anthropomorphic{} robot, was asking the participant to provide feedback.}
        \label{fig:exp_scene_humanoid}
    \end{subfigure}
    \hfill
    \begin{subfigure}[t]{.245\textwidth}
        \centering
        \includegraphics[width=\linewidth]{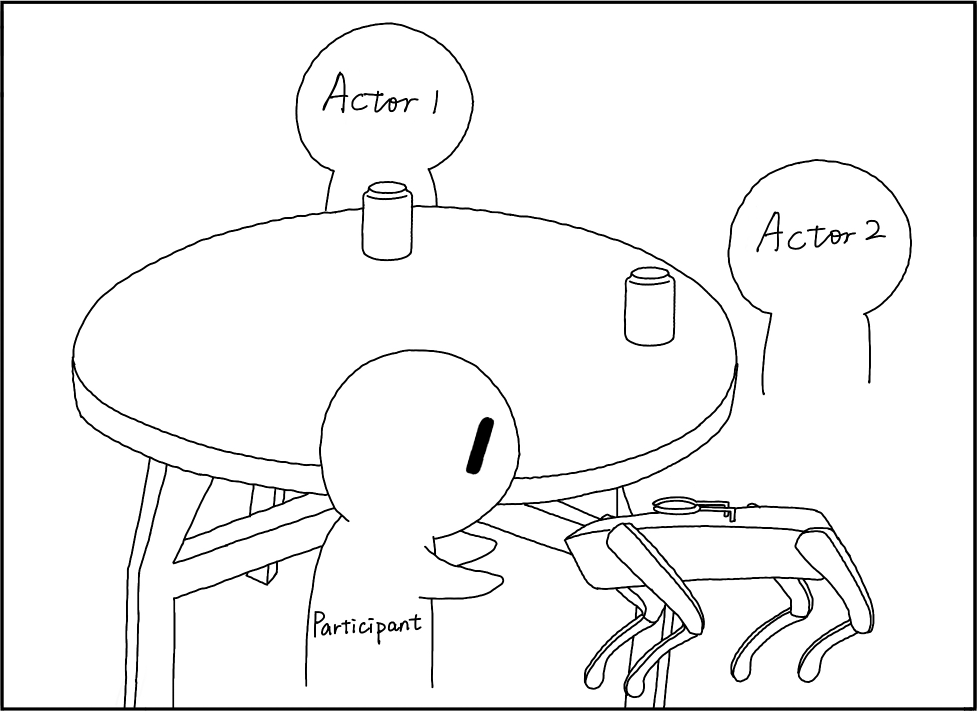}
        \medbreak
        \includegraphics[width=\linewidth]{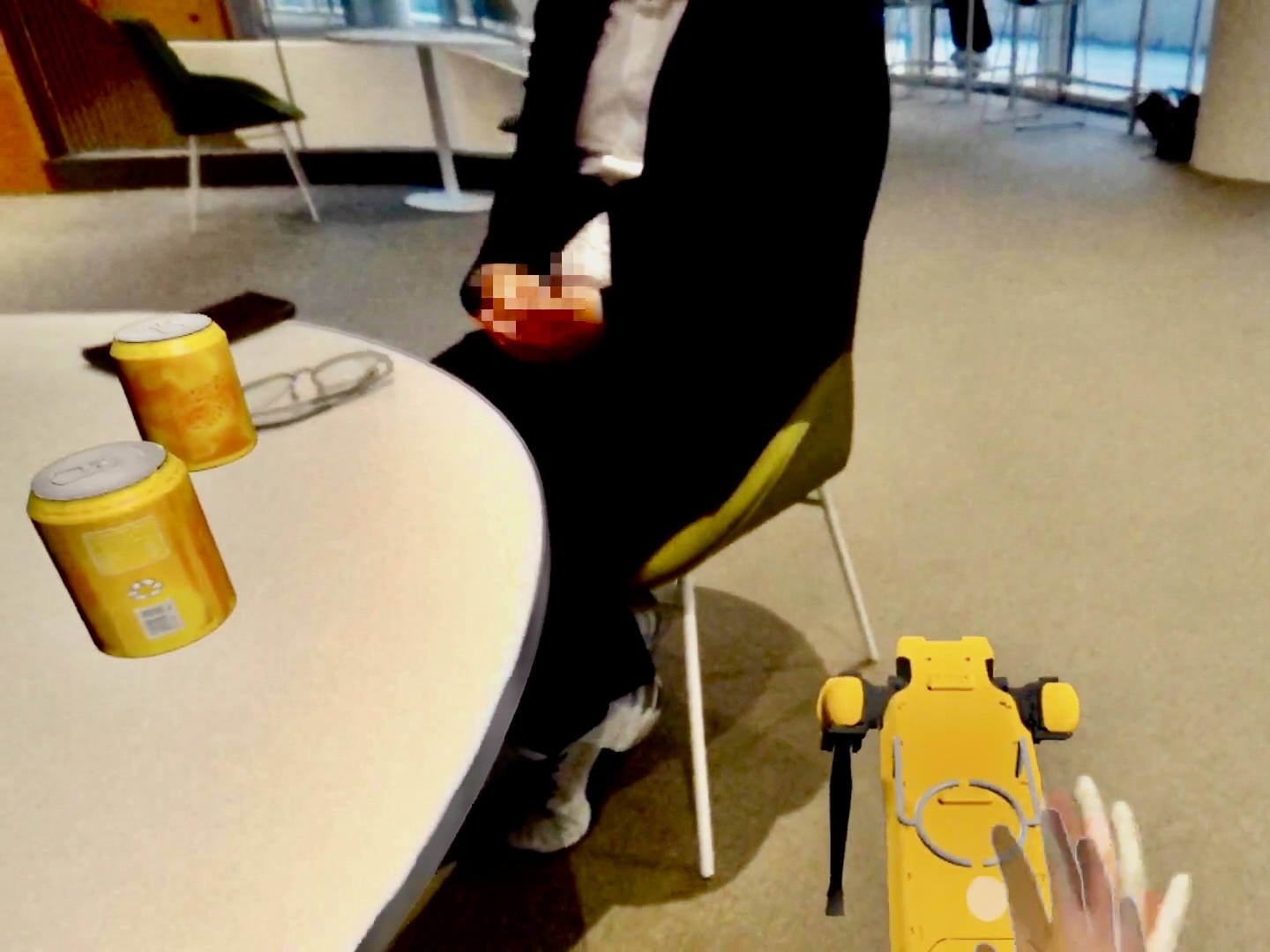}
        \subcaption{The participant was asking Spot, the \Zoomorphic{} robot, to go away.}
        \label{fig:exp_scene_quadruped}
    \end{subfigure}
    \hfill
    \begin{subfigure}[t]{.245\textwidth}
        \centering
        \includegraphics[width=\linewidth]{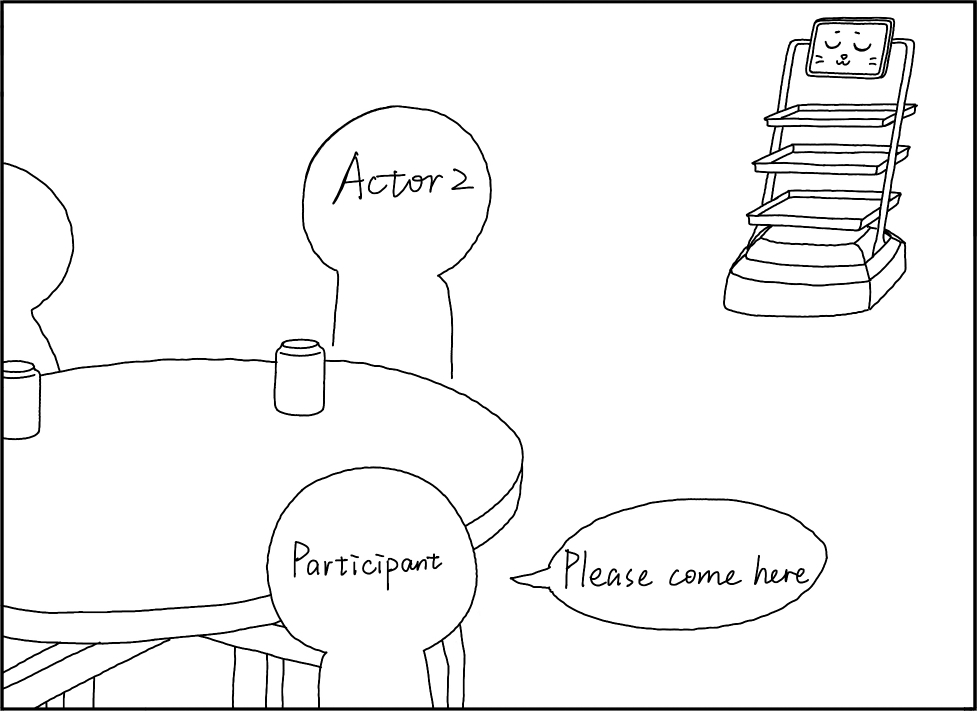}
        \medbreak
        \includegraphics[width=\linewidth]{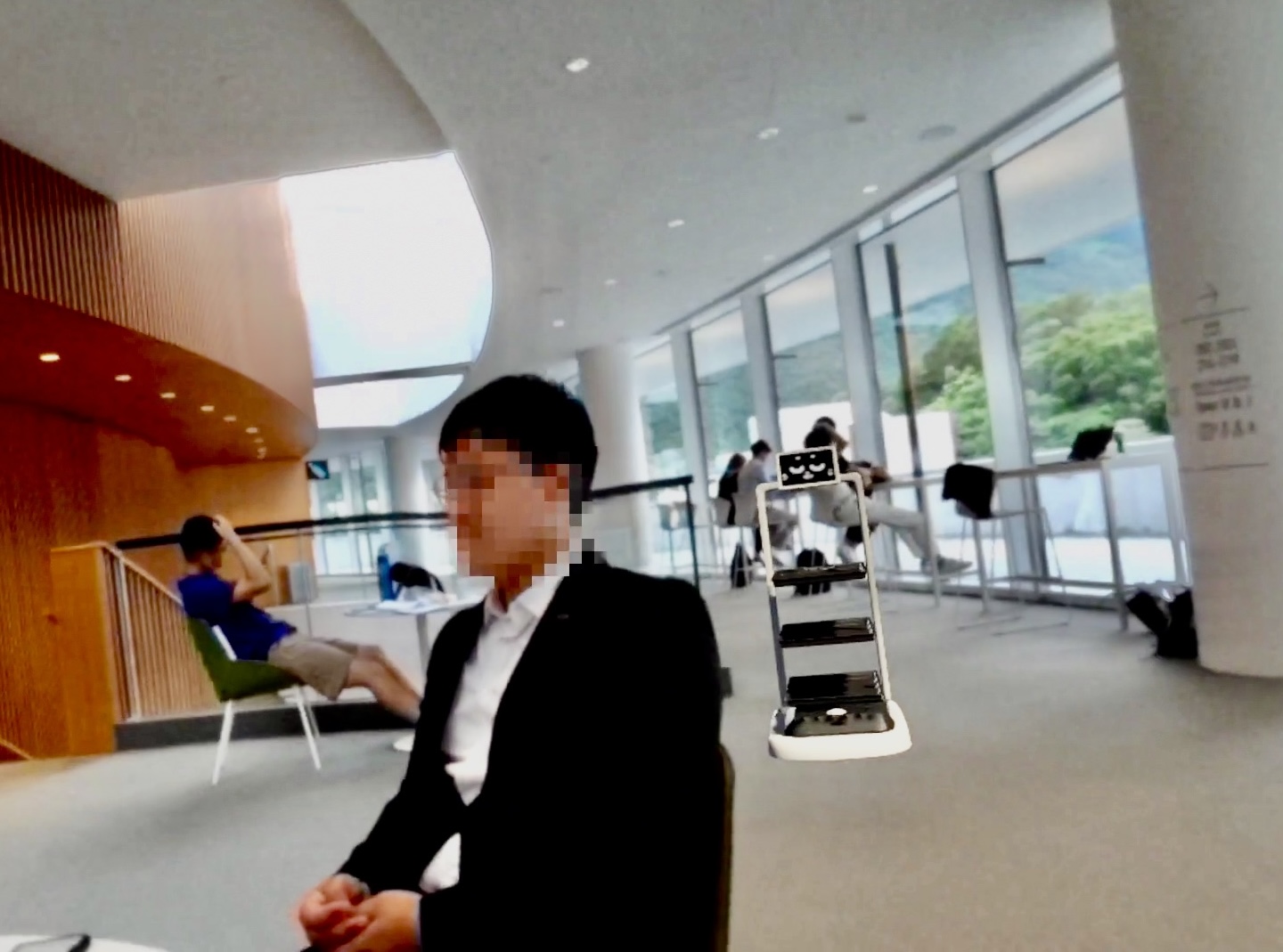}
        \subcaption{The participant was asking the \GroundedTechnical{} robot to come over.}
        \label{fig:exp_scene_waiter}
    \end{subfigure}
    \hfill
    \begin{subfigure}[t]{.245\textwidth}
        \centering
        \includegraphics[width=\linewidth]{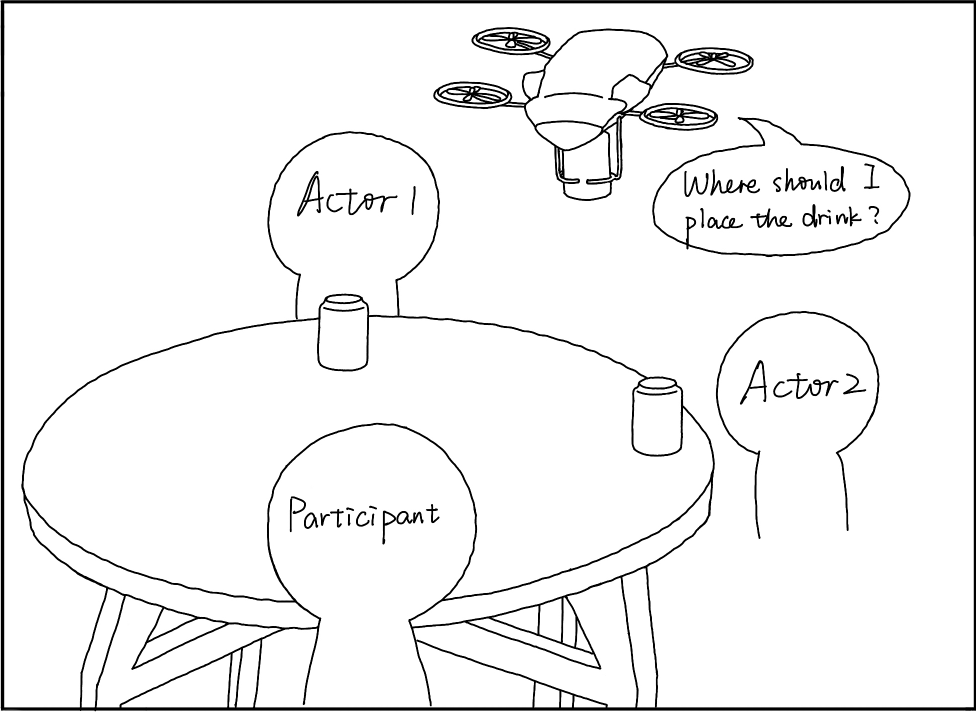}
        \medbreak
        \includegraphics[width=\linewidth]{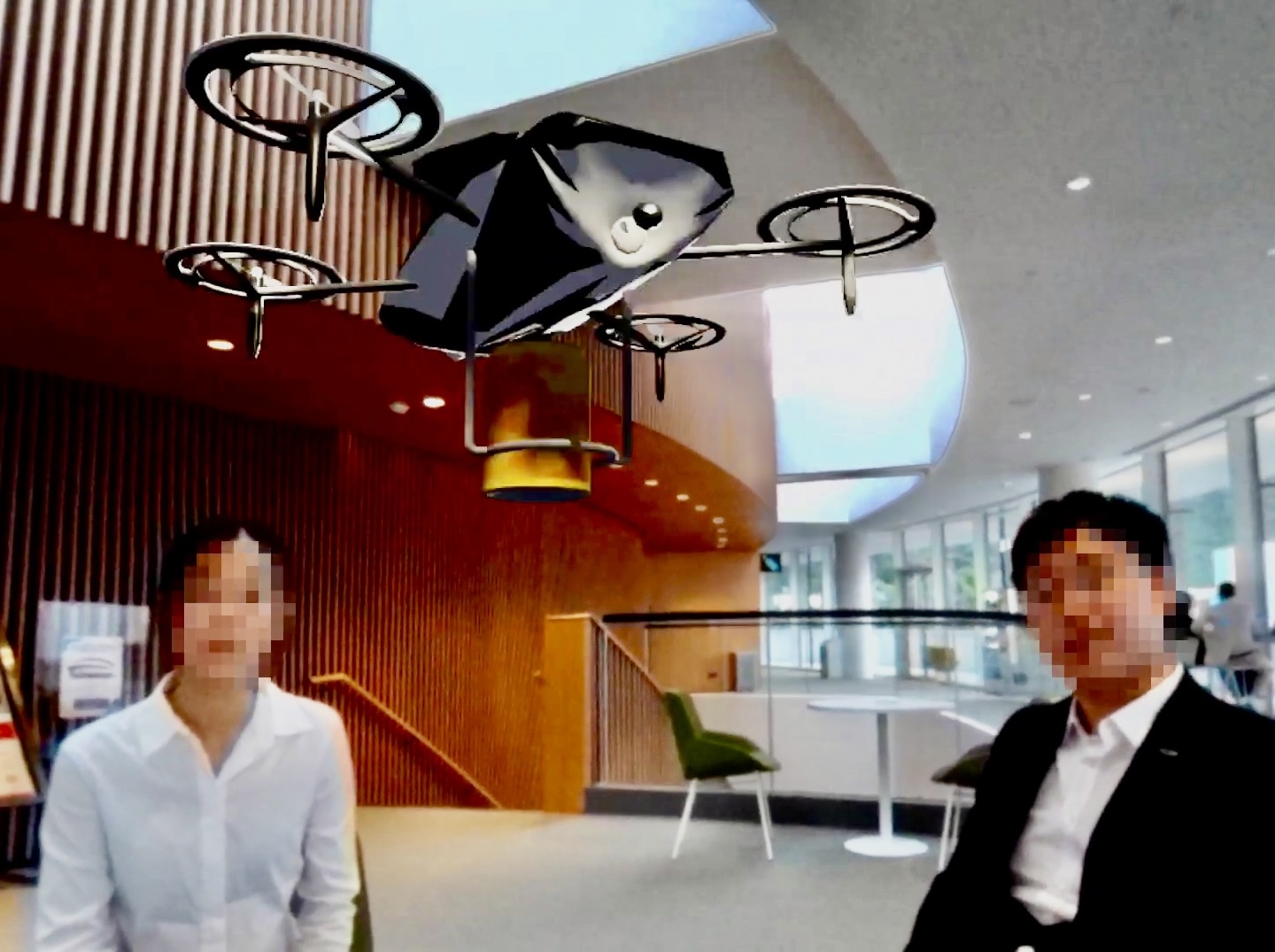}
        \subcaption{The \AerialTechnical{} robot carried the second cup of drink to the participant and asked him/her where to place it.}
        \label{fig:exp_scene_drone}
    \end{subfigure}
    \caption{Experiment scenes with four robot morphologies. First row: Scenario illustrations showing (a) \Anthropomorphic{}, (b) \Zoomorphic{}, (c) \GroundedTechnical{}, and (d) \AerialTechnical{} robots. Second row: Corresponding first-person participant viewpoints captured.}
    \Description{Experiment scenes with four robot morphologies, showing scenario illustrations (first row) and corresponding participant viewpoints (second row).  (a) Pepper, the \Anthropomorphic{} robot, was asking the participant to provide feedback. (b) The participant was asking Spot, the \Zoomorphic{} robot, to go away. (c) The participant was asking the \GroundedTechnical{} robot to come over. (d) The \AerialTechnical{} robot carried the second cup of drink to the participant and asked him/her where to place it. (e-h) Images captured from the participants' viewpoint during their interaction with each of the above robots.}
    \label{fig:illustration_of_interaction_with_four_robots}
\end{figure*}

\subsection{Elicitation Study}

\subsubsection{Participants}
\label{method:elicitation-study:participants}
We recruited $24$ participants ($7$ females and $17$ males) aged between $19$ to $33$ ($M = 24.08, SD = 2.7$). 
Most of them are students from various majors at local universities, which made them more relatable as future position seekers in our experiment.
We randomly assigned each participant to two robot groups, with $12$ participants in each group.
The participants' familiarities with the four robot morphologies are shown in \Cref{tab:participant-familiarity}.
We also collected data on the participants' dominant hand, with $23$ being right-handed and only one left-handed.
All participants were compensated at a rate of $12\$$ per hour.

\subsubsection{Experiment Settings}
In our elicitation study, the participants were asked to engage in a coffee chat with two potential employers, with a robot waiter serving around. The chat was conducted in a shared public workspace where people could occasionally pass by. 
The robots were simulated in the Augmented Reality (AR) of Quest Pro\footnote{\url{https://www.meta.com/quest/quest-pro/}} using Unity\footnote{version 2022.3.32f1, \url{https://unity.com/}} and controlled with Wizard of Oz (WoZ). We used Quest Pro's built-in cameras to capture participants' eye gazes, facial expressions, and body poses, and recorded a first-view video from Unity to visualize the eye gaze. We also set up another external camera to capture participants' whole bodies so that we can collect other modalities of their social cues. 
Two actors played the roles of potential employers (advisors) for future jobs, internships, or advanced study according to participants' status and goals.
To ensure the authenticity of the AR simulation, the actors relied on the sound effects to determine the ongoing task and the robot's state and acted accordingly to pretend that they could see the robot. \Cref{fig:illustration_of_interaction_with_four_robots} shows the figure illustrations of the experiment scenes and screenshots of Quest Pro Recordings from the participant's viewpoint.

\subsubsection{Design}

We proposed a mixed-design study, incorporating 
{two within-subjects variables, \role{} (nominal, two levels: \RoleS{} and \RoleL{}, as described in \Cref{method: simulation: social-encounter}) and \referent{} (nominal, $13$ levels, detailed in \Cref{tab:referents-used-in-our-experiment}),} and
one mixed-design variable, \robot{} (nominal, four levels: \Anthropomorphic{}, \Zoomorphic{}, \GroundedTechnical{}, \AerialTechnical{}, as introduced in Section~\ref{method:simulation:robot} and fig.~\ref{fig:robot-form-screenshot}).
During the coffee chat, participants engaged in both conversational roles across two sessions: 
1) the \RoleS{}, who was asked to introduce themselves and their experiences to the potential employers; 
2) the \RoleL{}, who mainly listened to the conversation between the two potential employers, but could also cut in to ask questions or make comments. 
Across these two sessions, participants interacted with two out of four robot morphologies (\AerialTechnical{}, \GroundedTechnical{}, \Anthropomorphic{}, \Zoomorphic{}), which acted as waiters in the study scenario.
Participants elicited interactions for all $13$ referents described in \Cref{method:simulation:referents}.
Each \role{} group included $24$ participants, while each \robot{} group comprised $12$ participants. This resulted in six participants for each combination of \role{} $\times$ \robot{}.
To reduce possible learning and order effects, the combination of robot and role, together with their order of presentation, was counterbalanced. \enlargethispage{12pt}

\subsubsection{Procedure}

Upon arrival, participants first filled out a demographic questionnaire and signed a consent form. 
Then, they were briefed on the requirements and the whole flow of the experiment:
Participants were {instructed} to {utilize} any social cues {they} deemed natural and appropriate {for conveying} a set of referents to the robot while {maintaining} their attention on the conversation as much as possible; 
Additionally, participants were asked to envision future robot waiters and assume that these robots are capable of identifying any modality of their social cues.
For each referent, participants were encouraged to elicit as many ways as possible, but this was not forced to ensure {that} they were not overloaded and {to} guarantee the naturalness of the elicited cues.
After confirming that the participants had a clear understanding of the task, participants were instructed to put on the Quest Pro headset, and calibrate the eye-tracking function within the headset. 
Before commencing the formal experiment, the participants were given preliminary exposure to all the referents through a robot demonstration to familiarize them with them and avoid omissions or misinterpretations. 
The participants then elicited social cues under their assigned conditions for the two sessions. 
After the experiment, the participants were presented with the synchronized videos from all four sources (three videos for eye gaze, facial expression, and body pose separately, and one video from the external camera for the full body) and were asked to identify the social cues they used for each referent using retrospective think-aloud. 
Finally, we conducted a semi-structured interview to learn 1) their rationales for choosing or not choosing specific modalities and social cues and 2) how their ways of interaction might be similar or different under different robot forms and roles.
The whole process took approximately $2$ hours, with around $30$ minutes for filling out the questionnaire and briefing, $10$ minutes for warm-up, $10$ minutes for each session, and $40$ minutes for retrospective think-aloud and interview. Breaks were guaranteed between each stage.

\subsection{Data Analysis}
\label{sec:method-data-analysis}
\subsubsection{Social Cue Coding}
We adopted an iterative approach to develop our codebook for social cues. Considering that most participants were only able to intuitively propose one social cue for each referent, we only coded the first elicited cue. Our social cue coding process involved the following steps:
\begin{itemize}
    \item \textbf{Step 1. Initial Coding}: We first reviewed all the videos to list the observed frequent patterns to form our initial codebook. The codes were roughly classified by different body parts, \eg arm and hand, eye, upper body, \etc
    \item \textbf{Step 2. Discussion and Iterative Development of the Codebook}: We used an iterative process to refine our codebook. We divided the whole dataset ($24$ participants $\times$ $2$ sessions $= 48$ sessions) into $6$ batches, each comprising $8$ sessions. Two coders first used the initial codebook to code one batch of data and then discussed how to resolve conflicts with the intervention of a third researcher. Codes may be added, removed, or reorganized during this process, and the revised codebook is used to code the next batch. After two batches, there are no more updates to the codebook. After the third batch, we used Perreault \& Leigh's approach (\textit{$I_r$}) \cite{perreaultReliabilityNominalData1989} to compute the inter-rater reliability (IRR) between the two coders, and got an IRR of $0.896$, exceeding the pre-set requirement of $0.7$, thus the iterative development of the codebook finished.
    \item \textbf{Step 3. Formal Coding}: The remaining data was divided up and coded independently by the two coders. The data used in the iterative development step was also re-coded using the finalized codebook.
\end{itemize}

\begin{figure*}[htbp]
    \centering
    \begin{subfigure}{.245\textwidth}
        \centering
        \includegraphics[width=\textwidth]{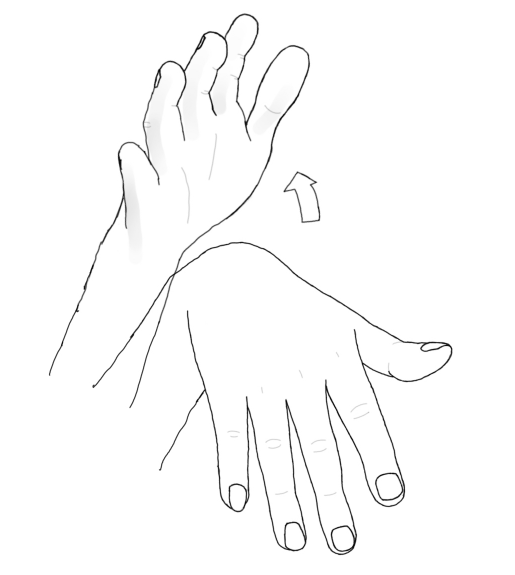}
        \subcaption{Dismissive Wave}
        \label{fig:top_cues_dismissive_wave}
    \end{subfigure}
    \hfil
    \begin{subfigure}{.245\textwidth}
        \centering
        \includegraphics[width=\textwidth]{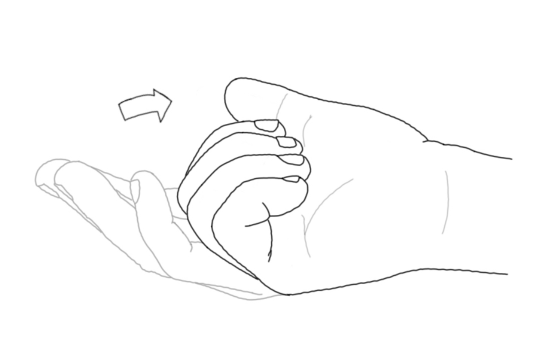}
        \subcaption{Beckoning Wave}
        \label{fig:top_cues_beckoning_wave}
    \end{subfigure}
    \hfil
    \begin{subfigure}{.245\textwidth}
        \centering
        \includegraphics[width=\textwidth]{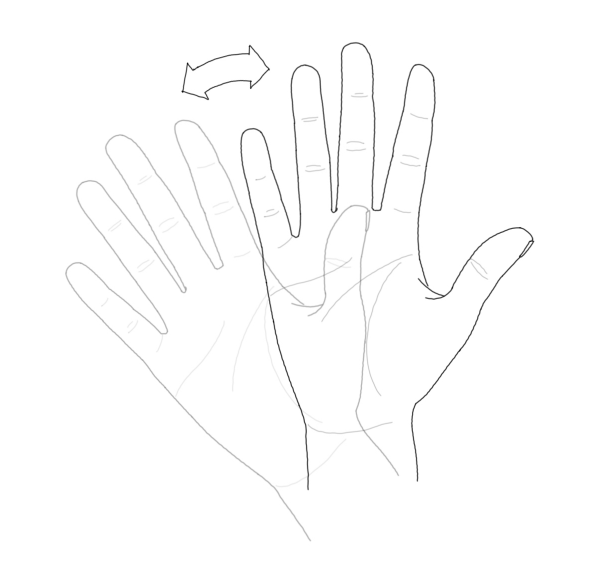}
        \subcaption{Wave}
        \label{fig:top_cues_wave}
    \end{subfigure}
    \hfil
    \begin{subfigure}{.245\textwidth}
        \centering
        \includegraphics[width=\textwidth]{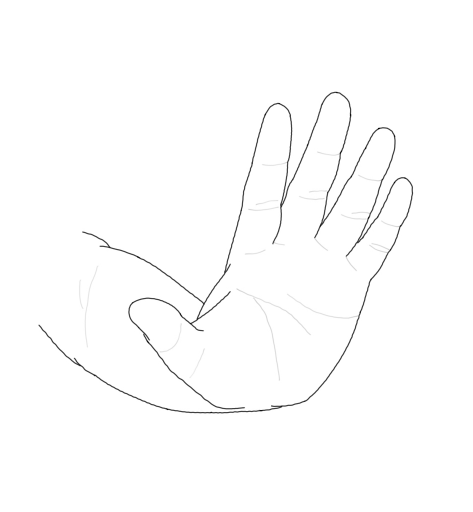}
        \subcaption{Palm to Stop the Robot}
        \label{fig:top_cues_palm_stop}
    \end{subfigure}
    \hfil
    \begin{subfigure}{.26\textwidth}
        \centering
        \includegraphics[width=\textwidth]{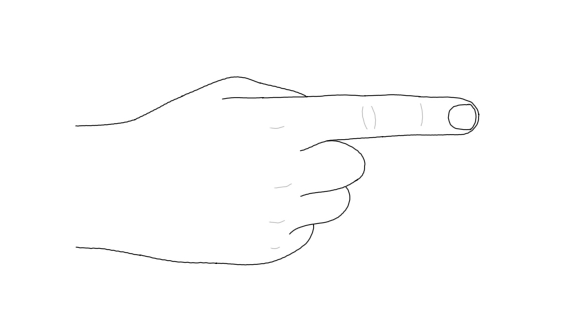}
        \subcaption{Finger Point (to Different Targets)}
        \label{fig:top_cues_finger_point}
    \end{subfigure}
    \hfil
    \begin{subfigure}{.48\textwidth}
        \centering
        \includegraphics[width=\textwidth]{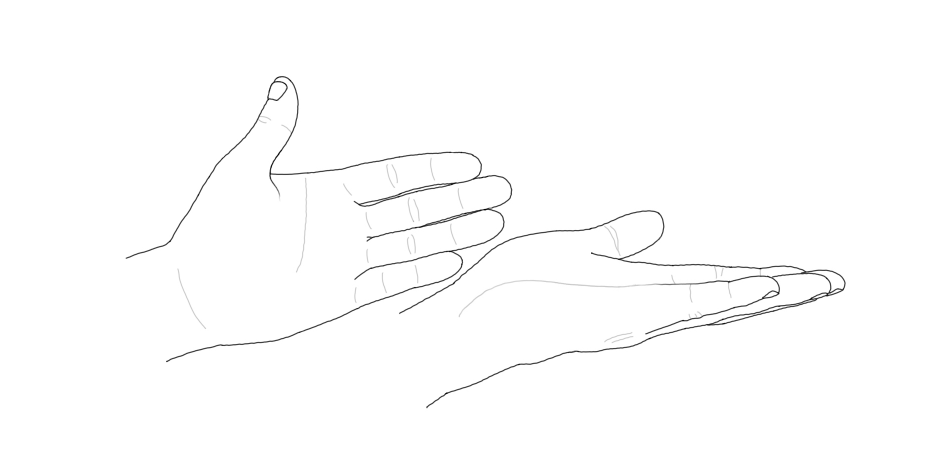}
        \subcaption{Palm Point (to Different Targets)}
        \label{fig:top_cues_palm_point}
    \end{subfigure}
    \hfil
    \begin{subfigure}{.245\textwidth}
        \centering
        \includegraphics[width=\textwidth]{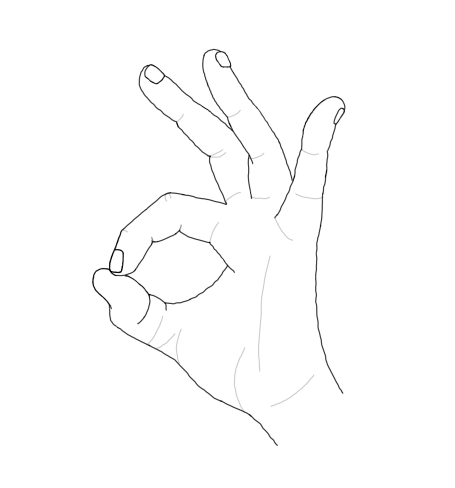}
        \subcaption{OK}
        \label{fig:top_cues_ok}
    \end{subfigure}
    \hfil
    \begin{subfigure}{.205\textwidth}
        \centering
        \includegraphics[width=\textwidth]{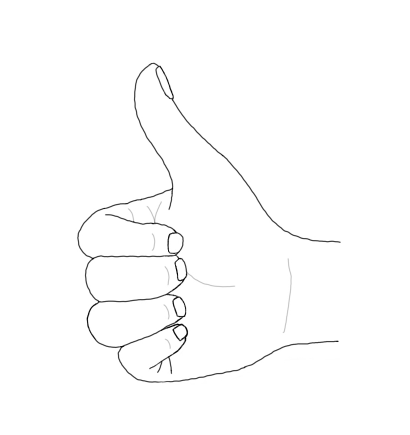}
        \subcaption{Thumb Up}
        \label{fig:top_cues_thumb_up}
    \end{subfigure}
    \hfil
    \begin{subfigure}{.205\textwidth}
        \centering
        \includegraphics[width=\textwidth]{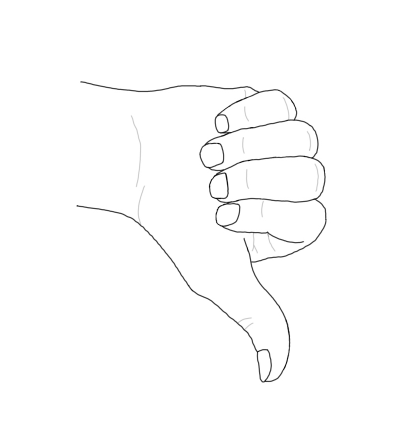}
        \subcaption{Thumb Down}
        \label{fig:top_cues_thumb_down}
    \end{subfigure}
    \hfil
    \begin{subfigure}{.265\textwidth}
        \centering
        \includegraphics[width=\textwidth]{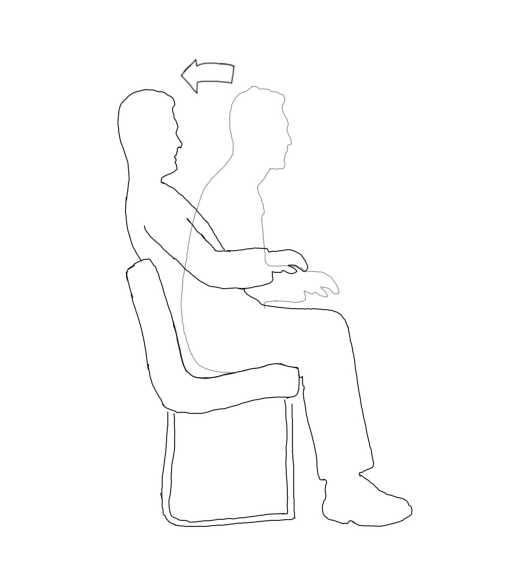}
        \subcaption{Lean Back}
        \label{fig:top_cues_lean_back}
    \end{subfigure}
    \hfil
    \begin{subfigure}{.305\textwidth}
        \centering
        \includegraphics[width=\textwidth]{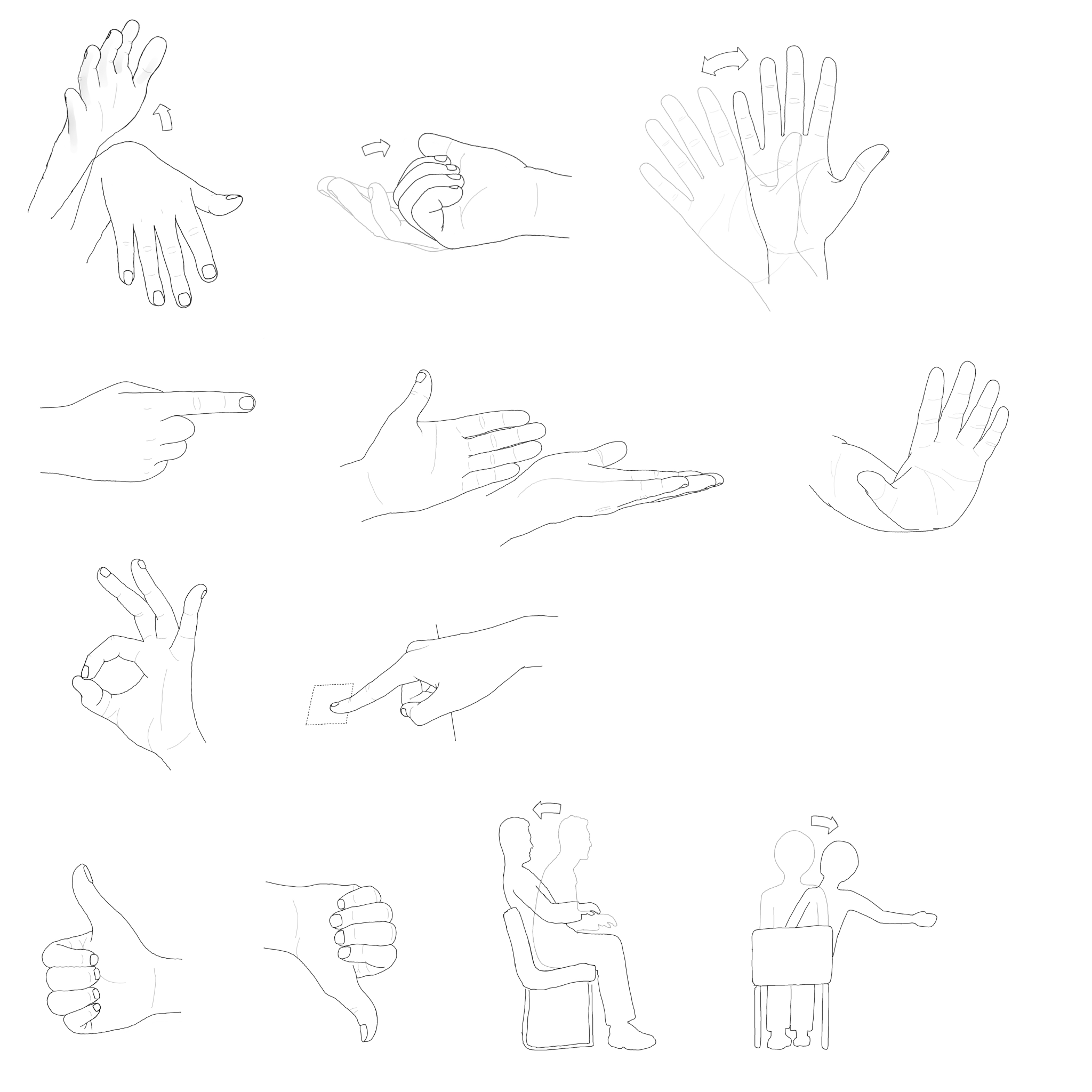}
        \subcaption{Lean Towards the Robot}
        \label{fig:top_cues_lean_right}
    \end{subfigure}
    \caption{The figure illustrations of some common codes in the \handAndArmGesture{} modality.}
    \Description{The figure illustrations of some common codes in the Gesture modality: (a) dismissive wave; (b) beckoning wave; (c) wave; (d) palm to stop the robot; (e) finger point (to different targets); (f) palm point (to different targets); (g) ok; (h) thumb up; (i) thumb down; and Upper Body Modality: (j) lean back; (k) lean towards the robot.}
    \label{fig:top_cues}
\end{figure*}

Upon finalizing the codebook, we obtained $85$ \textbf{codes}, categorized by different \textbf{modality (body parts)}, including \handAndArmGesture{}, \verbal{}, \eyeGaze{}, \headMotion{}, \etc{}
For each modality, there are \textbf{articulation codes} which decompose the \textbf{social cues} elicited by participants into basic elements.
Articulations are coded as explicit or implicit, where explicit signals refer to those explicitly reported by participants during the post-interaction retrospective think-aloud process, and implicit signals are those observed but not verbally acknowledged by participants.
Due to the nuances of \handAndArmGesture{} and \verbal{}, we further developed \textbf{feature codes} describing their detailed characteristics.
A complete description of our codebook is as follows. Some of the most frequently used codes are illustrated in \Cref{fig:top_cues}.

\begin{itemize}
    \item \handAndArmGesture{}
    \begin{itemize}
        \item Articulation codes for \handAndArmGesture{} are unique gesture names, including waving gestures (\eg \GNI{}, \GNII{}, \GNIII{}), pointing gestures (\eg \GPPII{}, \GPPIII{}, \GFPII{}) and others (\eg \GNIV{}, \GNXIII{}, \GNVII{}). Note that multiple gestures may be used in one social cue, so each code is treated as a binary variable.
        \item Feature codes of \handAndArmGesture{} further describe characteristics that apply to any gesture. This includes \handedness{} (near-side hand/far-side hand), \numberOfHands{} (single hand/both hands), \handHeight{} (lower than table level/not lower than table but lower than head level/head level/above head level), \repetition{} (repetition/no repetition).
    \end{itemize}
    
    \item \verbal{}
    \begin{itemize}
        \item Articulation codes for \verbal{} include the \exactContent{} and \speechAct{} \cite{sadockSpeechActDistinctions1988} of the content (declarative, interrogative, imperative, exclamative, or short interjections).
        \item Feature codes for \verbal{} include \volume{} (decrease/no change/increase, compared with the chatting volume), \VPronoun{} (exist/not exist), and \politeness{}. To measure \politeness{}, we adopted the features in \texttt{politeness} R package \footnote{https://cran.r-project.org/web/packages/politeness/vignettes/politeness.html\#politeness-features} which is built upon previous research on computational linguistics \cite{brownPolitenessUniversalsLanguage1987, danescu-niculescu-mizilComputationalApproachPoliteness2013,voigtLanguagePoliceBody2017}, and manually selected $9$ features that are applicable in HRI scenario as our codes: \politenessApology{}, \politenessCouldYou{}, \politenessFirstPersonPlural{}, \politenessFormalTitle{}, \politenessGratitude{}, \politenessHello{}, \politenessPlease{}, \politenessPositiveEmotion{}, \politenessReasoning{}, and \politenessSubjectivity{}.
    \end{itemize}

    \item \eyeGaze{}. Articulation codes for \eyeGaze{} are defined based on the gaze target: \EGI{}, \EGII{} (further divided into \EGIIW{}, \EGIIC{}), \EGIII{} and \EGIV{}.

    \item \headMotion{}. Articulation codes for \headMotion{} are divided into two parts: 1) head motions that are caused by eye gaze, including \HGI{}, \HGII{} (further divided into \HGIIW{}, \HGIIC{}), \HGIII{}, \HGIV{}. 2) Head motions that are irrelevant to gaze, including \HMI{}, \HMII{}, \HMIII{} and \HMIV{}.

    \item \upperBody{}. Articulation codes for \upperBody{} include \BI{}, \BII{}, \BIII{}.

    \item \legAndFoot{}. Articulation codes for \legAndFoot{} include \LFI{}, \LFII{}, \LFIII{}, \LFIV{}.

    \item \facialExpression{}. Articulation codes for \facialExpression{} include \FEI{}, \FEII{}, \FEIII{} and \FEIV{}.

\end{itemize}

For subsequent analysis, we define a \textbf{unique social cue} as a unique set of articulations from all modalities in our codes.

\subsubsection{Statistics}
Our study was a mixed design with two within-subjects variables (\role{} and \referent{}) and one mixed-design variable (\robot{}).
The dependent variables are the modalities (\modality) and the feature codes (\handAndArmGesture{} and \verbal{}) of human social cues.
We analyzed the \referent{} variable with the statistical distributions and the agreement rates,
and fitted Cumulative Link Mixed Models to assess the effects of \role{} and \robot{} on the \modality{}, \handAndArmGesture{} and \verbal{} features of human social cues.

\paragraph{Agreement Rate (AR)}
To understand participants' consensus on each \referent{} representing their intentions, we calculated agreement rates for all unique cues. 
As introduced in \cite{vatavuFormalizingAgreementAnalysis2015, muehlhausNeedThirdArm2023}, the agreement rate for each referent $r$ is calculated with the following function:
\begin{equation}
    AR(r) = \frac{\lvert{P}\rvert}{\lvert{P}\rvert - 1}\sum_{P_i \subseteq P} \left(\frac{\lvert{P_i}\rvert}{P}\right)^2 - \frac{1}{\lvert{P}\rvert - 1},
\end{equation}
where $P$ is the set of all social cues elicited for the referent $r$, and $P_i$ is the $i^{th}$ subset of identical codes in $P$.
The margins for interpretation are $ \leq 0.1$ for low agreement, $0.1 < AR \leq 0.3$ for medium agreement, $0.3 < AR \leq 0.5$ for high agreement, and $AR > 0.5$ for very high agreement \cite{vatavuFormalizingAgreementAnalysis2015}.

\paragraph{Statistical Analysis}

We employed Cumulative Link Mixed Models fitted with the adaptive Gauss-Hermite quadrature approximation to assess the effects of \robot{} and \role{} \cite{christensenOrdinalRegressionModels2024} on the \modality{}, \handAndArmGesture{} and \verbal{} features of human social cues.
Following the practice of \cite{andreiTakeSeatMake2024}, we treated participants and \referent{} as a random effect since we regard the referents as samples from the common human-robot interactions in the coffee chat scenario.
The baseline for the \role{} variable was set to \RoleL{}.
Since the location coefficients may vary in p-values for different baselines chosen for the \robot{} variable, we expanded \robot{} to four dummy variables \AerialTechnical{}, \GroundedTechnical{}, \Anthropomorphic{}, and \Zoomorphic{} with levels $0,1$ to compare the effects of each robot morphology.
The following cumulative link mixed model was fitted to the dependent variable $Y$ (\modality{}, \handAndArmGesture{} or \verbal{} features of human social cues):
\begin{equation}
    \begin{aligned}
        \text{logit}(P(Y_i \leq j))
         & = \theta_j + \beta_1{(\role{}_i)}                                 \\
         & + \beta_2(\AerialTechnical{}_i) + \beta_3(\GroundedTechnical{}_i) \\
         & + \beta_4(\Anthropomorphic{}_i) + \beta_5(\Zoomorphic{}_i)        \\
         & + \gamma_1(\role{}_i \times \AerialTechnical{}_i)                 \\
         & + \gamma_2(\role{}_i \times \GroundedTechnical{}_i)               \\
         & + \gamma_3(\role{}_i \times \Anthropomorphic{}_i)                 \\
         & + \gamma_4(\role{}_i \times \Zoomorphic{}_i)                      \\
         & + u(\referent{}_i) + v(\text{participant}_i),
    \end{aligned}
\end{equation}
where $i = 1,\ldots,78$, and $j$ represents the ordinal level of the dependent variable $Y$.
To analyze the statistical significance of each model term, we applied forward selection with likelihood-ratio chi-squared tests \cite{christensenOrdinalRegressionModels2024} for each pair of models with progressive complexities.

\subsection{Interview Data Analysis}
All the video-recorded interviews are transcribed into text. 
Three researchers first independently read and watched the videos of the retrospective think-aloud and the interview process and familiarized themselves with $12$ out of $24$ experiments. 
Alongside the analysis, each researcher also referred to the corresponding videos where particular social cues were used. 
The whole research group then identified major themes focusing on participants' rationales in choosing their social cues through rounds of discussions. 
Due to the qualitative nature of the data, we did not conduct an inter-rater reliability check~\cite{McDonald2019}. 
Instead, we ensured the reliability of the analysis through both independent coding and cross-checking among the research team.

\section{Result}

\subsection{Overall Statistics of Social Cues}
\label{sec:result-overall}

\begin{figure*}[htbp]
    \centering
    \includegraphics[width=1\textwidth]{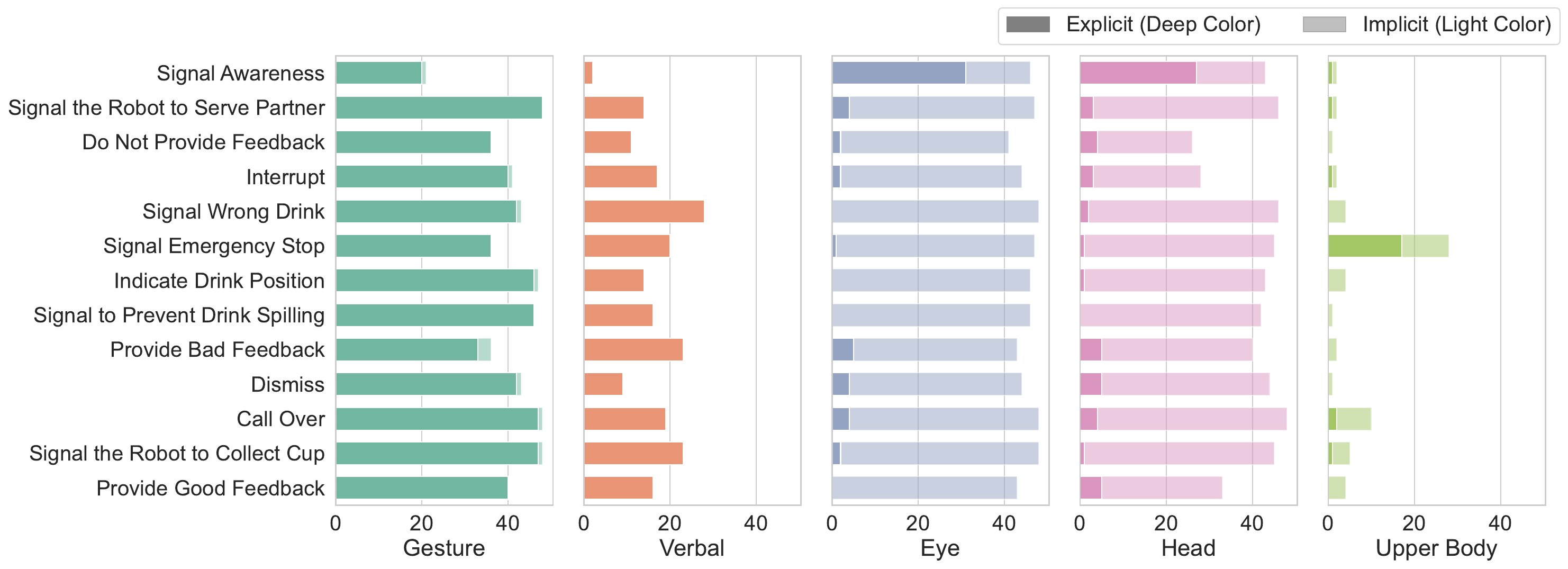}
    \caption{Explicit (dark) and implicit (light) \modality{} usages by \referent.}
    \Description{Explicit (dark) and implicit (light) modality usages by referents. Modalities include hand/arm gestures, verbal cues, eye gaze, head motion, and upper body movements. Interaction contexts encompass signaling robot actions (e.g., serving, emergency stop), feedback, and interventions. Color shading indicates deliberate (explicit) versus incidental (implicit) signal usage. Overall, the most frequently used modalities are eye gaze (94.71\%), gesture (85.42\%), and head motion (84.78\%). Verbal (66.03\%) and upper body (89.42\%) follow, while leg and foot (1.12\%), as well as facial expression (0.64\%), are not depicted in the figure due to their relatively low frequency of use. The most frequently used explicit modalities are gesture (83.81\%) and verbal (66.03\%), while the most frequently used implicit modalities are eye gaze (86.38\%) and head motion (76.44\%). When checking by referents, we can observe that Signal awareness usually involves frequent usage of explicit eye gaze (31/48) and explicit head motion (27/48), with few usages of verbal (2/48). Signaling an emergency stop usually involves significantly more usage of the upper body (17/48) modality. Besides, Signal the Wrong Drink shows more verbal usage (28/48), possibly because this referent inherently carries more complex semantics.}
    \label{fig:modality_by_referent}
\end{figure*}

We analyzed $624$ observed (explicit $+$ implicit) social cues ($= 24$ participants $\times$ $2$ sessions $\times 13$ referents) from participants' first choice to signal the service robot. 
We coded $3,387$ modality articulations, $3,318$ gestural features, and $823$ verbal features. 
\Cref{fig:modality_by_referent} presents the distributions of \modality{} by each referent. 
Overall, the most frequently used modalities are \eyeGaze{} ($94.71\%$), \handAndArmGesture{} ($85.42\%$), and \headMotion{} ($84.78\%$). \verbal{} ({$33.97\%$}) and \upperBody{} ({$10.58\%$}) follow, while \legAndFoot{} ($1.12\%$) as well as \facialExpression{} ($0.64\%$) are not depicted in the figure due to their relatively low frequency of use.
The most frequently used explicit modalities are \handAndArmGesture{} ($83.81\%$) and \verbal{} ({$33.97\%$}), while the most frequently used implicit modalities are \eyeGaze{} ($86.38\%$) and \headMotion{} ($76.44\%$). 
When checking by referents, we can observe that \ReferentoneI{} usually involves frequent usage of explicit \eyeGaze{} ($31/48$) and explicit \headMotion{} ($27/48$), with few usages of \verbal{} ($2/48$). 
\ReferenttwoII{} usually involves significantly more usage of \upperBody{} ($17/48$) modality. 
Besides, \ReferenttwoI{} shows more usage of \verbal{} ($28/48$), possibly because this referent inherently carries more complex semantics.

We counted the number of unique social cues elicited by participants and listed the calculated agreement rates ($AR(r)$) by each referent $r$ for both observed cues and explicit cues, as shown in \Cref{tab:top_cues_by_referents}.
The agreement rates range from $0.02$ (low agreement, $AR \leq 0.1$) to $0.23$  (medium agreement, $0.1 < AR \leq 0.3$) for observed cues and from $0.03$ (low agreement, $AR \leq 0.1$) to $0.30$ (medium agreement, $0.1 < AR \leq 0.3$) for explicit cues.
The rather low agreement rates suggest the potential effects of the other two IVs and the complexity of expressing the corresponding intentions through social cues, given the diverse combinations of articulations.
Despite the low agreement rates, we still observed patterns of social cue articulations from the top $3$ cues for each referent in Table~\ref{tab:top_cues_by_referents}.
Referent \ReferentoneI{} frequently includes head gaze and beckoning waves (Figure~\ref{fig:top_cues_beckoning_wave}). 
Referent \ReferentoneII{} emphasizes pointing gestures towards the table's opposite side, where most pointing gestures were pointing with an open palm (Figure~\ref{fig:top_cues_palm_point}) to show politeness. 
Referents \ReferentoneIII{} and \ReferentoneIV{} often feature \dismissiveWave{} (Figure~\ref{fig:top_cues_dismissive_wave}) or \wave{} (Figure~\ref{fig:top_cues_wave}) gestures to {express ``no'' and dismissive intention to the robot waiter.}
Referent \ReferenttwoI{} involves \wave{} to say ``wrong'' with additional gestural or verbal explanations. 
For referent \ReferenttwoII{}, participants use \palmStop{} (Figure~\ref{fig:top_cues_palm_stop}), often with verbal imperatives. 
Referent \ReferenttwoIII{} highlights pointing and tapping for drink placement. 
Referent \ReferenttwoIV{} shows participants often \GNXIII{}, sometimes with verbal cues, showing that participants would like to correct the drink themselves rather than asking the robot to do so.
Participants use \thumbDown{} (Figure~\ref{fig:top_cues_thumb_down}) and declarative words to express negative sentiments when they \ReferenttwoV{}. 
The gesture \dismissiveWave{} is common for the referent \ReferentthreeI{}, 
and the referent \ReferentthreeII{} involves more \beckoningWave{}, 
which are aligned with the referent meanings. 
Referent \ReferentthreeIII{} relies on pointing to the opposite side, 
and \ReferentthreeIV{} combines \thumbUp{} (Figure~\ref{fig:top_cues_thumb_up}) gestures with verbal declarative.

\subsection{Effects of \robot{} and \role{} On Social Cues}
\label{sec:result-effects}
Due to the large number of dependent variables from social cue codes, we only present significant results from the statistical analysis of the effects on independent variables \robot{} or \role{} on different modalities, modality articulations, and main modality features in the following sections.

\subsubsection{\modality}
No significant effect of \role{} or \robot{} on the use of \handAndArmGesture{} or \headMotion{} was found in the mixed model analysis, and since the labels are rare for \legAndFoot{} and \facialExpression{}, we did not include them in the analysis. The effects of \role{} and \robot{} on the use of \verbal{}, \eyeGaze{} and \upperBody{} are summarized in Table~\ref{tab:regression-result-modality} and presented below.

\paragraph{\verbal}
\label{sec:result-modality-verbal}
During the coffee chat, participants chose to talk to the robot waiter among $25.96\%$ of their social cues in the \RoleL{} session, and participants chose to use verbal among $41.99\%$ of their social cues in the \RoleS{} session.
The LR test for the mixed models shows a significant effect on \role{} ($LR=40.091, p < 0.001$) and \AerialTechnical{} ($LR=4.445, p<0.05$) respectively, and the interaction effect of \role{} and \AerialTechnical{} is also significant ($LR=9.316, p<0.01$).
The positive and significant coefficients of \RoleS{} ($1.137, p < 0.01$) and \RoleS{} $\times$ \AerialTechnical{} ($2.855, p < 0.01$) suggest that participants are more likely to include verbal cues in the \RoleS{} session than in the \RoleL{} session.
The interaction effect of \RoleS{} $\times$ \AerialTechnical{} ($2.855, p < 0.01$) indicates that the \AerialTechnical{} robot amplifies verbal cue usage only in the \RoleS{} session: participants used verbal cues $56.4\%$ of the time with the \AerialTechnical{} robot versus $37.2\%$ with non-\AerialTechnical{} robots. 

\aptLtoX[graphic=no,type=html]{\begin{table*}
    \caption{Top observed and self-reported cues by referents, {where $AR(r)$ stands for the agreement rate of the referent}.} \label{tab:top_cues_by_referents}
\begin{tabular}[c]{m{135pt}m{50pt}m{135pt}m{50pt}m{135pt}}
    \toprule
    \multirow{2.3}{*}{{\textbf{Referents}}}
     &
    \multicolumn{2}{c}{\textbf{Observed (Explicit + Implicit)}}
     &
    \multicolumn{2}{c}{\textbf{Self-Reported (Explicit)}}
    \\
    \cmidrule{2-5}
     &
    {$AR(r)$}
     &
    Top 3 Cues
     &
    {$AR(r)$}
     &
    Top 3 Cues
    \\
    \midrule    
    \multirow{5.8}{*}{{\ReferentoneI}}
     &
    {\multirow{5.8}{*}{ $0.19$ }}
     &
    Head gaze turn to robot ($33\%$)
     &
    { \multirow{5.8}{*}{ $0.18$ } }
     &
    Head gaze turn to robot ($35\%$)
    \\
    \arrayrulecolor{gray}%
    \cmidrule{3-3}%
    \cmidrule{5-5}%
    \arrayrulecolor{black}%
     &
    {}
     &
    Head gaze turn to robot + beckoning wave ($27\%$)
     &
    {}
     &
    Beckoning wave ($21\%$)
    \\
    \arrayrulecolor{gray}%
    \cmidrule{3-3}%
    \cmidrule{5-5}%
    \arrayrulecolor{black}%
     &
    {}
     &
    Head gaze turn to robot + nod ($8\%$)
     &
    {}
     &
    Head gaze turn to robot + beckoning wave ($10\%$)
    \\
    \midrule
    \multirow{7.8}{*}{\parbox{120pt}{\ReferentoneII}}
     &
    {\multirow{7.8}{*}{ $0.06$ }}
     &
    Head gaze turn to robot and the correct drink + palm point to the opposite side of the table ($23\%$)
     &
    {\multirow{7.8}{*}{ $0.09$ } }
     &
    Palm point to the opposite side of the table ($27\%$)
    \\
    \arrayrulecolor{gray}%
    \cmidrule{3-3}%
    \cmidrule{5-5}%
    \arrayrulecolor{black}%
     &
    {}
     &
    Head gaze turn to robot and the correct drink + verbal (imperative) + palm point to the opposite side of the table ($6\%$)
     &
    {}
     &
    Palm point from robot to the opposite side of the table ($12\%$)
    \\
    \arrayrulecolor{gray}%
    \cmidrule{3-3}%
    \cmidrule{5-5}%
    \arrayrulecolor{black}%
     &
    {}
     &
    Head gaze turn to robot and the correct drink + palm point from robot to the opposite side of the table ($6\%$)
     &
    {}
     &
    Finger point to the opposite side of the table ($10\%$)
    \\
    \midrule
    \multirow{4.8}{*}{{\ReferentoneIII}}
     &
    {\multirow{4.8}{*}{ $0.03$ }}
     &
    Glance at robot + wave hand ($10\%$)
     &
    { \multirow{4.8}{*}{ $0.07$ } }
     &
    Dismissive wave ($17\%$)
    \\
    \arrayrulecolor{gray}%
    \cmidrule{3-3}%
    \cmidrule{5-5}%
    \arrayrulecolor{black}%
     &
    {}
     &
    Head gaze turn to robot + dismissive wave ($10\%$)
     &
    {}
     &
    Wave hand ($15\%$)
    \\
    \arrayrulecolor{gray}%
    \cmidrule{3-3}%
    \cmidrule{5-5}%
    \arrayrulecolor{black}%
     &
    {}
     &
    Glance at robot ($8\%$)
     &
    {}
     &
    (Do nothing) ($10\%$)
    \\
    \midrule
    \multirow{5.8}{*}{{\ReferentoneIV}}
     &
    {\multirow{5.8}{*}{ $0.02$ }}
     &
    Head gaze turn to robot + dismissive wave ($13\%$)
     &
    { \multirow{5.8}{*}{ $0.05$ } }
     &
    Dismissive wave ($15\%$)
    \\
    \arrayrulecolor{gray}%
    \cmidrule{3-3}%
    \cmidrule{5-5}%
    \arrayrulecolor{black}%
     &
    {}
     &
    Head gaze turn to robot + show palm to stop the robot ($8\%$)
     &
    {}
     &
    Show palm to stop the robot ($13\%$)
    \\
    \arrayrulecolor{gray}%
    \cmidrule{3-3}%
    \cmidrule{5-5}%
    \arrayrulecolor{black}%
     &
    {}
     &
    Head gaze turn to robot + wave hand + verbal (declarative) ($6\%$)
     &
    {}
     &
    Wave hand + verbal (declarative) ($8\%$)
    \\
    \midrule
    \multirow{5.8}{*}{{\ReferenttwoI}}
     &
    {\multirow{5.8}{*}{ $0.02$ }}
     &
    Head gaze turn to robot and the wrong drink + wave hand + dismissive wave ($10\%$)
     &
    { \multirow{5.8}{*}{ $0.03$ } }
     &
    Wave hand ($10\%$)
    \\
    \arrayrulecolor{gray}%
    \cmidrule{3-3}%
    \cmidrule{5-5}%
    \arrayrulecolor{black}%
     &
    {}
     &
    Head gaze turn to robot and wrong drink + verbal (declarative) ($8\%$)
     &
    {}
     &
    Wave hand + dismissive wave ($10\%$)
    \\
    \arrayrulecolor{gray}%
    \cmidrule{3-3}%
    \cmidrule{5-5}%
    \arrayrulecolor{black}%
     &
    {}
     &
    Head gaze turn to robot and wrong drink + wave hand + verbal (declarative) ($6\%$)
     &
    {}
     &
    Wave hand + verbal (declarative) ($8\%$)
    \\
    \midrule
    \multirow{5.8}{*}{{\ReferenttwoII}}
     &
    {\multirow{5.8}{*}{ $0.04$ }}
     &
    Head gaze turn to robot + show palm to stop the robot ($19\%$)
     &
    { \multirow{5.8}{*}{ $0.07$ } }
     &
    Show palm to stop the robot ($21\%$)
    \\
    \arrayrulecolor{gray}%
    \cmidrule{3-3}%
    \cmidrule{5-5}%
    \arrayrulecolor{black}%
     &
    {}
     &
    Head gaze turn to robot + lean away from the robot ($8\%$)
     &
    {}
     &
    Show palm to stop the robot + verbal (imperative) ($15\%$)
    \\
    \arrayrulecolor{gray}%
    \cmidrule{3-3}%
    \cmidrule{5-5}%
    \arrayrulecolor{black}%
     &
    {}
     &
    Head gaze turn to robot + show palm to stop the robot + verbal (imperative) ($6\%$)
     &
    {}
     &
    Show palm to stop the robot + verbal (declarative) ($6\%$)
    \\
    \midrule
    \multirow{7.8}{*}{{\ReferenttwoIII}}
     &
    {\multirow{7.8}{*}{ $0.03$ }}
     &
    Head gaze turn to robot and the correct drink + palm point to the place for the drink + tap the table ($15\%$)
     &
    { \multirow{7.8}{*}{ $0.11$ } }
     &
    Palm point to the place for the drink + tap the table ($23\%$)
    \\
    \arrayrulecolor{gray}%
    \cmidrule{3-3}%
    \cmidrule{5-5}%
    \arrayrulecolor{black}%
     &
    {}
     &
    Head gaze turn to robot and the correct drink + head gaze turn to and palm point to the place for the drink ($6\%$)
     &
    {}
     &
    Palm point to the place for the drink ($23\%$)
    \\
    \arrayrulecolor{gray}%
    \cmidrule{3-3}%
    \cmidrule{5-5}%
    \arrayrulecolor{black}%
     &
    {}
     &
    Head gaze turn to the robot and the correct drink + palm point to the place for the drink ($6\%$)
     &
    {}
     &
    Finger point to the place for the drink + tap the table + verbal (imperative) ($6\%$)
    \\
    \midrule
    \multirow{6.9}{*}{ \parbox{120pt}{\ReferenttwoIV} }
     &
    {\multirow{6.9}{*}{ $0.17$ }}
     &
    Head gaze turn to robot and the correct drink + hold the drink with hand ($42\%$)
     &
    { \multirow{6.9}{*}{ $0.30$ } }
     &
    Hold the drink with hand ($54\%$)
    \\
    \arrayrulecolor{gray}%
    \cmidrule{3-3}%
    \cmidrule{5-5}%
    \arrayrulecolor{black}%
     &
    {}
     &
    Head gaze turn to robot and the correct drink + hold the drink with hand + verbal (imperative) ($6\%$)
     &
    {}
     &
    Hold the drink with hand + verbal (short exclamation)($6\%$)
    \\
    \arrayrulecolor{gray}%
    \cmidrule{3-3}%
    \cmidrule{5-5}%
    \arrayrulecolor{black}%
     &
    {}
     &
    Head gaze turn to robot and the correct drink + hold the drink with hand + verbal (declarative) ($6\%$)
     &
    {}
     &
    Hold the drink with hand + verbal (imperative) ($6\%$)
    \\
    \midrule
    \multirow{4.8}{*}{{\ReferenttwoV}}
     &
    {\multirow{4.8}{*}{ $0.04$ }}
     &
    Head gaze turn to robot + thumb down ($15\%$)
     &
    { \multirow{4.8}{*}{ $0.07$ } }
     &
    verbal (declarative) ($19\%$)
    \\
    \arrayrulecolor{gray}%
    \cmidrule{3-3}%
    \cmidrule{5-5}%
    \arrayrulecolor{black}%
     &
    {}
     &
    Head gaze turn to robot + verbal (declarative) ($8\%$)
     &
    {}
     &
    Thumb down ($17\%$)
    \\
    \arrayrulecolor{gray}%
    \cmidrule{3-3}%
    \cmidrule{5-5}%
    \arrayrulecolor{black}%
     &
    {}
     &
    Head gaze turn to robot + wave hand ($6\%$)
     &
    {}
     &
    Wave hand + dismissive wave ($6\%$)
    \\
    \midrule
    \multirow{5.8}{*}{\ReferentthreeI}
     &
    {\multirow{5.8}{*}{ $0.23$ }}
     &
    Head gaze turn to robot + dismissive wave ($48\%$)
     &
    { \multirow{5.8}{*}{ $0.20$ } }
     &
    Dismissive wave ($44\%$)
    \\
    \arrayrulecolor{gray}%
    \cmidrule{3-3}%
    \cmidrule{5-5}%
    \arrayrulecolor{black}%
     &
    {}
     &
    Head gaze turn to robot + show palm to stop the robot ($6\%$)
     &
    {}
     &
    Show palm to stop the robot ($8\%$)
    \\
    \arrayrulecolor{gray}%
    \cmidrule{3-3}%
    \cmidrule{5-5}%
    \arrayrulecolor{black}%
     &
    {}
     &
    (Do nothing) ($4\%$)
     &
    {}
     &
    Head gaze turn to robot + dismissive wave ($6\%$)
    \\
    \midrule
    \multirow{5.8}{*}{{\ReferentthreeII}}
     &
    {\multirow{5.8}{*}{ $0.16$ }}
     &
    Head gaze turn to robot + beckoning wave ($40\%$)
     &
    { \multirow{5.8}{*}{ $0.19$ } }
     &
    Beckoning wave ($42\%$)
    \\
    \arrayrulecolor{gray}%
    \cmidrule{3-3}%
    \cmidrule{5-5}%
    \arrayrulecolor{black}%
     &
    {}
     &
    Head gaze turn to robot + beckoning wave + verbal (imperative) ($8\%$)
     &
    {}
     &
    Beckoning wave + verbal (imperative) ($13\%$)
    \\
    \arrayrulecolor{gray}%
    \cmidrule{3-3}%
    \cmidrule{5-5}%
    \arrayrulecolor{black}%
     &
    {}
     &
    Head gaze turn to robot + beckoning wave + lean back ($4\%$)
     &
    {}
     &
    Beckoning wave + verbal (declarative) ($6\%$)
    \\
    \midrule
    \multirow{6.9}{*}{ \parbox{120pt}{\ReferentthreeIII} }
     &
    {\multirow{6.9}{*}{ $0.02$ }}
     &
    Head gaze turn to robot + palm point to the opposite side of the table + verbal (imperative) ($10\%$)
     &
    { \multirow{6.9}{*}{ $0.04$ } }
     &
    Palm point to the opposite side of the table ($15\%$)
    \\
    \arrayrulecolor{gray}%
    \cmidrule{3-3}%
    \cmidrule{5-5}%
    \arrayrulecolor{black}%
     &
    {}
     &
    Head gaze turn to robot + palm point to the opposite side of the table ($8\%$)
     &
    {}
     &
    Palm point to the opposite side of the table + verbal (imperative) ($13\%$)
    \\
    \arrayrulecolor{gray}%
    \cmidrule{3-3}%
    \cmidrule{5-5}%
    \arrayrulecolor{black}%
     &
    {}
     &
    Head gaze turn to robot + head gaze turn to the opposite side of the table + palm point to the opposite side of the table ($8\%$)
     &
    {}
     &
    Finger point to the opposite side of the table ($8\%$)
    \\
    \midrule
    \multirow{4.8}{*}{{\ReferentthreeIV}}
     &
    {\multirow{4.8}{*}{ $0.10$ }}
     &
    Head gaze turn to robot + thumb up ($23\%$)
     &
    { \multirow{4.8}{*}{ $0.26$ } }
     &
    Thumb up ($46\%$)
    \\
    \arrayrulecolor{gray}%
    \cmidrule{3-3}%
    \cmidrule{5-5}%
    \arrayrulecolor{black}%
     &
    {}
     &
    Head gaze turn to robot + thumb up + verbal (declarative) ($17\%$)
     &
    {}
     &
    Thumb up + verbal (declarative) ($21\%$)
    \\
    \arrayrulecolor{gray}%
    \cmidrule{3-3}%
    \cmidrule{5-5}%
    \arrayrulecolor{black}%
     &
    {}
     &
    Glance at robot + thumb up ($15\%$)
     &
    {}
     &
    verbal (declarative) ($10\%$)
    \\ \bottomrule
     \end{tabular}
\end{table*}}{
\onecolumn
\setlength{\colonewidth}{\dimexpr .2\textwidth - 2\tabcolsep}
\setlength{\coltwowidth}{\dimexpr .07\textwidth - 2\tabcolsep}
\setlength{\colthreewidth}{\dimexpr .33\textwidth - 2\tabcolsep}
\begin{longtable}[c]{m{\colonewidth}m{\coltwowidth}m{\colthreewidth}m{\coltwowidth}m{\colthreewidth}}
    \caption{Top observed and self-reported cues by referents, where $AR(r)$ stands for the agreement rate of the referent.} \label{tab:top_cues_by_referents}
    \\
    \toprule
    \multirow{2.3}{*}{{\textbf{Referents}}}
     &
    \multicolumn{2}{c}{\textbf{Observed (Explicit + Implicit)}}
     &
    \multicolumn{2}{c}{\textbf{Self-Reported (Explicit)}}
    \\
    \cmidrule{2-5}
     &
    {$AR(r)$}
     &
    Top 3 Cues
     &
    {$AR(r)$}
     &
    Top 3 Cues
    \\
    \midrule
    \endfirsthead

    \multicolumn{5}{c}%
    {Table \thetable.\ Top cues by referents, where $AR(r)$ stands for the agreement rate of the referent.\ (continued)}
    \\
    \midrule
    \multirow{2.3}{*}{\textbf{Referents}}
     &
    \multicolumn{2}{c}{\textbf{Observed (Explicit + Implicit)}}
     &
    \multicolumn{2}{c}{\textbf{Self-Reported (Explicit)}}
    \\
    \cmidrule{2-5}
     &
    {$AR(r)$}
     &
    Top 3 Cues
     &
    {$AR(r)$}
     &
    Top 3 Cues
    \\
    \midrule
    \endhead

    \multirow{5.8}{*}{{\ReferentoneI}}
     &
    {\multirow{5.8}{*}{ $0.19$ }}
     &
    Head gaze turn to robot ($33\%$)
     &
    { \multirow{5.8}{*}{ $0.18$ } }
     &
    Head gaze turn to robot ($35\%$)
    \\
    \arrayrulecolor{gray}%
    \cmidrule{3-3}%
    \cmidrule{5-5}%
    \arrayrulecolor{black}%
     &
    {}
     &
    Head gaze turn to robot + beckoning wave ($27\%$)
     &
    {}
     &
    Beckoning wave ($21\%$)
    \\
    \arrayrulecolor{gray}%
    \cmidrule{3-3}%
    \cmidrule{5-5}%
    \arrayrulecolor{black}%
     &
    {}
     &
    Head gaze turn to robot + nod ($8\%$)
     &
    {}
     &
    Head gaze turn to robot + beckoning wave ($10\%$)
    \\
    \midrule
    \multirow{7.8}{*}{\parbox{\colonewidth}{\ReferentoneII}}
     &
    {\multirow{7.8}{*}{ $0.06$ }}
     &
    Head gaze turn to robot and the correct drink + palm point to the opposite side of the table ($23\%$)
     &
    {\multirow{7.8}{*}{ $0.09$ } }
     &
    Palm point to the opposite side of the table ($27\%$)
    \\
    \arrayrulecolor{gray}%
    \cmidrule{3-3}%
    \cmidrule{5-5}%
    \arrayrulecolor{black}%
     &
    {}
     &
    Head gaze turn to robot and the correct drink + verbal (imperative) + palm point to the opposite side of the table ($6\%$)
     &
    {}
     &
    Palm point from robot to the opposite side of the table ($12\%$)
    \\
    \arrayrulecolor{gray}%
    \cmidrule{3-3}%
    \cmidrule{5-5}%
    \arrayrulecolor{black}%
     &
    {}
     &
    Head gaze turn to robot and the correct drink + palm point from robot to the opposite side of the table ($6\%$)
     &
    {}
     &
    Finger point to the opposite side of the table ($10\%$)
    \\
    \midrule
    \multirow{4.8}{*}{{\ReferentoneIII}}
     &
    {\multirow{4.8}{*}{ $0.03$ }}
     &
    Glance at robot + wave hand ($10\%$)
     &
    { \multirow{4.8}{*}{ $0.07$ } }
     &
    Dismissive wave ($17\%$)
    \\
    \arrayrulecolor{gray}%
    \cmidrule{3-3}%
    \cmidrule{5-5}%
    \arrayrulecolor{black}%
     &
    {}
     &
    Head gaze turn to robot + dismissive wave ($10\%$)
     &
    {}
     &
    Wave hand ($15\%$)
    \\
    \arrayrulecolor{gray}%
    \cmidrule{3-3}%
    \cmidrule{5-5}%
    \arrayrulecolor{black}%
     &
    {}
     &
    Glance at robot ($8\%$)
     &
    {}
     &
    (Do nothing) ($10\%$)
    \\
    \midrule
    \multirow{5.8}{*}{{\ReferentoneIV}}
     &
    {\multirow{5.8}{*}{ $0.02$ }}
     &
    Head gaze turn to robot + dismissive wave ($13\%$)
     &
    { \multirow{5.8}{*}{ $0.05$ } }
     &
    Dismissive wave ($15\%$)
    \\
    \arrayrulecolor{gray}%
    \cmidrule{3-3}%
    \cmidrule{5-5}%
    \arrayrulecolor{black}%
     &
    {}
     &
    Head gaze turn to robot + show palm to stop the robot ($8\%$)
     &
    {}
     &
    Show palm to stop the robot ($13\%$)
    \\
    \arrayrulecolor{gray}%
    \cmidrule{3-3}%
    \cmidrule{5-5}%
    \arrayrulecolor{black}%
     &
    {}
     &
    Head gaze turn to robot + wave hand + verbal (declarative) ($6\%$)
     &
    {}
     &
    Wave hand + verbal (declarative) ($8\%$)
    \\
    \midrule
    \multirow{5.8}{*}{{\ReferenttwoI}}
     &
    {\multirow{5.8}{*}{ $0.02$ }}
     &
    Head gaze turn to robot and the wrong drink + wave hand + dismissive wave ($10\%$)
     &
    { \multirow{5.8}{*}{ $0.03$ } }
     &
    Wave hand ($10\%$)
    \\
    \arrayrulecolor{gray}%
    \cmidrule{3-3}%
    \cmidrule{5-5}%
    \arrayrulecolor{black}%
     &
    {}
     &
    Head gaze turn to robot and wrong drink + verbal (declarative) ($8\%$)
     &
    {}
     &
    Wave hand + dismissive wave ($10\%$)
    \\
    \arrayrulecolor{gray}%
    \cmidrule{3-3}%
    \cmidrule{5-5}%
    \arrayrulecolor{black}%
     &
    {}
     &
    Head gaze turn to robot and wrong drink + wave hand + verbal (declarative) ($6\%$)
     &
    {}
     &
    Wave hand + verbal (declarative) ($8\%$)
    \\
    \midrule
    \multirow{5.8}{*}{{\ReferenttwoII}}
     &
    {\multirow{5.8}{*}{ $0.04$ }}
     &
    Head gaze turn to robot + show palm to stop the robot ($19\%$)
     &
    { \multirow{5.8}{*}{ $0.07$ } }
     &
    Show palm to stop the robot ($21\%$)
    \\
    \arrayrulecolor{gray}%
    \cmidrule{3-3}%
    \cmidrule{5-5}%
    \arrayrulecolor{black}%
     &
    {}
     &
    Head gaze turn to robot + lean away from the robot ($8\%$)
     &
    {}
     &
    Show palm to stop the robot + verbal (imperative) ($15\%$)
    \\
    \arrayrulecolor{gray}%
    \cmidrule{3-3}%
    \cmidrule{5-5}%
    \arrayrulecolor{black}%
     &
    {}
     &
    Head gaze turn to robot + show palm to stop the robot + verbal (imperative) ($6\%$)
     &
    {}
     &
    Show palm to stop the robot + verbal (declarative) ($6\%$)
    \\
    \midrule
    \multirow{7.8}{*}{{\ReferenttwoIII}}
     &
    {\multirow{7.8}{*}{ $0.03$ }}
     &
    Head gaze turn to robot and the correct drink + palm point to the place for the drink + tap the table ($15\%$)
     &
    { \multirow{7.8}{*}{ $0.11$ } }
     &
    Palm point to the place for the drink + tap the table ($23\%$)
    \\
    \arrayrulecolor{gray}%
    \cmidrule{3-3}%
    \cmidrule{5-5}%
    \arrayrulecolor{black}%
     &
    {}
     &
    Head gaze turn to robot and the correct drink + head gaze turn to and palm point to the place for the drink ($6\%$)
     &
    {}
     &
    Palm point to the place for the drink ($23\%$)
    \\
    \arrayrulecolor{gray}%
    \cmidrule{3-3}%
    \cmidrule{5-5}%
    \arrayrulecolor{black}%
     &
    {}
     &
    Head gaze turn to the robot and the correct drink + palm point to the place for the drink ($6\%$)
     &
    {}
     &
    Finger point to the place for the drink + tap the table + verbal (imperative) ($6\%$)
    \\
    \midrule
    \multirow{6.9}{*}{ \parbox{\colonewidth}{\ReferenttwoIV} }
     &
    {\multirow{6.9}{*}{ $0.17$ }}
     &
    Head gaze turn to robot and the correct drink + hold the drink with hand ($42\%$)
     &
    { \multirow{6.9}{*}{ $0.30$ } }
     &
    Hold the drink with hand ($54\%$)
    \\
    \arrayrulecolor{gray}%
    \cmidrule{3-3}%
    \cmidrule{5-5}%
    \arrayrulecolor{black}%
     &
    {}
     &
    Head gaze turn to robot and the correct drink + hold the drink with hand + verbal (imperative) ($6\%$)
     &
    {}
     &
    Hold the drink with hand + verbal (short exclamation)($6\%$)
    \\
    \arrayrulecolor{gray}%
    \cmidrule{3-3}%
    \cmidrule{5-5}%
    \arrayrulecolor{black}%
     &
    {}
     &
    Head gaze turn to robot and the correct drink + hold the drink with hand + verbal (declarative) ($6\%$)
     &
    {}
     &
    Hold the drink with hand + verbal (imperative) ($6\%$)
    \\
    \midrule
    \multirow{4.8}{*}{{\ReferenttwoV}}
     &
    {\multirow{4.8}{*}{ $0.04$ }}
     &
    Head gaze turn to robot + thumb down ($15\%$)
     &
    { \multirow{4.8}{*}{ $0.07$ } }
     &
    verbal (declarative) ($19\%$)
    \\
    \arrayrulecolor{gray}%
    \cmidrule{3-3}%
    \cmidrule{5-5}%
    \arrayrulecolor{black}%
     &
    {}
     &
    Head gaze turn to robot + verbal (declarative) ($8\%$)
     &
    {}
     &
    Thumb down ($17\%$)
    \\
    \arrayrulecolor{gray}%
    \cmidrule{3-3}%
    \cmidrule{5-5}%
    \arrayrulecolor{black}%
     &
    {}
     &
    Head gaze turn to robot + wave hand ($6\%$)
     &
    {}
     &
    Wave hand + dismissive wave ($6\%$)
    \\
    \midrule
    \multirow{5.8}{*}{\ReferentthreeI}
     &
    {\multirow{5.8}{*}{ $0.23$ }}
     &
    Head gaze turn to robot + dismissive wave ($48\%$)
     &
    { \multirow{5.8}{*}{ $0.20$ } }
     &
    Dismissive wave ($44\%$)
    \\
    \arrayrulecolor{gray}%
    \cmidrule{3-3}%
    \cmidrule{5-5}%
    \arrayrulecolor{black}%
     &
    {}
     &
    Head gaze turn to robot + show palm to stop the robot ($6\%$)
     &
    {}
     &
    Show palm to stop the robot ($8\%$)
    \\
    \arrayrulecolor{gray}%
    \cmidrule{3-3}%
    \cmidrule{5-5}%
    \arrayrulecolor{black}%
     &
    {}
     &
    (Do nothing) ($4\%$)
     &
    {}
     &
    Head gaze turn to robot + dismissive wave ($6\%$)
    \\
    \midrule
    \multirow{5.8}{*}{{\ReferentthreeII}}
     &
    {\multirow{5.8}{*}{ $0.16$ }}
     &
    Head gaze turn to robot + beckoning wave ($40\%$)
     &
    { \multirow{5.8}{*}{ $0.19$ } }
     &
    Beckoning wave ($42\%$)
    \\
    \arrayrulecolor{gray}%
    \cmidrule{3-3}%
    \cmidrule{5-5}%
    \arrayrulecolor{black}%
     &
    {}
     &
    Head gaze turn to robot + beckoning wave + verbal (imperative) ($8\%$)
     &
    {}
     &
    Beckoning wave + verbal (imperative) ($13\%$)
    \\
    \arrayrulecolor{gray}%
    \cmidrule{3-3}%
    \cmidrule{5-5}%
    \arrayrulecolor{black}%
     &
    {}
     &
    Head gaze turn to robot + beckoning wave + lean back ($4\%$)
     &
    {}
     &
    Beckoning wave + verbal (declarative) ($6\%$)
    \\
    \midrule
    \multirow{6.9}{*}{ \parbox{\colonewidth}{\ReferentthreeIII} }
     &
    {\multirow{6.9}{*}{ $0.02$ }}
     &
    Head gaze turn to robot + palm point to the opposite side of the table + verbal (imperative) ($10\%$)
     &
    { \multirow{6.9}{*}{ $0.04$ } }
     &
    Palm point to the opposite side of the table ($15\%$)
    \\
    \arrayrulecolor{gray}%
    \cmidrule{3-3}%
    \cmidrule{5-5}%
    \arrayrulecolor{black}%
     &
    {}
     &
    Head gaze turn to robot + palm point to the opposite side of the table ($8\%$)
     &
    {}
     &
    Palm point to the opposite side of the table + verbal (imperative) ($13\%$)
    \\
    \arrayrulecolor{gray}%
    \cmidrule{3-3}%
    \cmidrule{5-5}%
    \arrayrulecolor{black}%
     &
    {}
     &
    Head gaze turn to robot + head gaze turn to the opposite side of the table + palm point to the opposite side of the table ($8\%$)
     &
    {}
     &
    Finger point to the opposite side of the table ($8\%$)
    \\
    \midrule
    \multirow{4.8}{*}{{\ReferentthreeIV}}
     &
    {\multirow{4.8}{*}{ $0.10$ }}
     &
    Head gaze turn to robot + thumb up ($23\%$)
     &
    { \multirow{4.8}{*}{ $0.26$ } }
     &
    Thumb up ($46\%$)
    \\
    \arrayrulecolor{gray}%
    \cmidrule{3-3}%
    \cmidrule{5-5}%
    \arrayrulecolor{black}%
     &
    {}
     &
    Head gaze turn to robot + thumb up + verbal (declarative) ($17\%$)
     &
    {}
     &
    Thumb up + verbal (declarative) ($21\%$)
    \\
    \arrayrulecolor{gray}%
    \cmidrule{3-3}%
    \cmidrule{5-5}%
    \arrayrulecolor{black}%
     &
    {}
     &
    Glance at robot + thumb up ($15\%$)
     &
    {}
     &
    verbal (declarative) ($10\%$)
    \\ \bottomrule
    % \end{tabular}
\end{longtable}}

\twocolumn

\begin{table*}[htbp]
  \centering
  \caption{Regression coefficients for predicting \modality{} features using Cumulative Link Mixed Models with forward progressive selection using likelihood-ratio Chi-squared tests. For each factor, the baseline is indicated in parentheses. In the table, $***$: $p<0.001$; $**$: $p<0.01$; $*$: $p<0.05$. ``LR'' stands for likelihood-ratio for model selection. Non-significant interaction terms are not shown.}
  \label{tab:regression-result-modality}
  \begin{tabular}{ccclll}
    \toprule
    \multicolumn{3}{c}{{\modality{}}}
     & \verbal
     & \eyeGaze
     & \upperBody~
    \\
    \midrule
    \multirow{2.3}{*}{\role}
     &
    \multirow{2.3}{*}{
      \begin{tabular}[c]{c}
        \RoleS{}
        \\ (\RoleL{})
      \end{tabular}}
     &
    LR
     &
    \textbf{40.091***}
     &
    0.554
     &
    2.693
    \\
    \arrayrulecolor{gray}
    \cmidrule{3-6}
    \arrayrulecolor{black}
     &
     & coef
     & \textbf{1.137**}   & - & -
    \\
    \cmidrule{1-6}
    \multirow{10.5}{*}{\robot}
     &
    \multirow{2.3}{*}{
      \begin{tabular}[c]{c}
        \AerialTechnical
        \\
        (non-\AerialTechnical)
      \end{tabular}}
     &
    LR
     &
    \textbf{4.445*}
     &
    1.261
     &
    0.742
    \\
    \arrayrulecolor{gray}
    \cmidrule{3-6}
    \arrayrulecolor{black}
     &
     & coef
     & \textbf{-0.708}
     & -
     & -
    \\
    \cmidrule{2-6}
     &
    \multirow{2.3}{*}{
      \begin{tabular}[c]{c}
        \Anthropomorphic
        \\
        (non-\Anthropomorphic)
      \end{tabular}}
     &
    LR
     &
    2.998
     &
    1.369
     &
    0.268
    \\
    \arrayrulecolor{gray}
    \cmidrule{3-6}
    \arrayrulecolor{black}
     &
     & coef
     & -
     & -
     & -
    \\
    \cmidrule{2-6}
     &
    \multirow{2.3}{*}{
      \begin{tabular}[c]{c}
        \GroundedTechnical
        \\
        (non-\GroundedTechnical)
      \end{tabular}}
     &
    LR
     &
    0.463
     &
    0.002
     &
    0.166
    \\
    \arrayrulecolor{gray}
    \cmidrule{3-6}
    \arrayrulecolor{black}
     &
     & coef
     & -
     & -
     & -
    \\
    \cmidrule{2-6}
     &
    \multirow{2.3}{*}{
      \begin{tabular}[c]{c}\Zoomorphic
        \\
        (non-\Zoomorphic)
      \end{tabular}}
     &
    LR
     &
    1.280
     &
    \textbf{18.212***}
     &
    \textbf{34.596***}
    \\ \arrayrulecolor{gray}
    \cmidrule{3-6}
    \arrayrulecolor{black}
     &
     & coef
     & -
     & \textbf{-1.808***}
     & \textbf{2.170***}
    \\ \cmidrule{1-6}
    \multirow{2.3}{*}{
      \begin{tabular}[c]{c}
        \role{}
        \\
        $\times$ \robot{}
      \end{tabular}}
    % \multirow{2.3}{*}{\role{} $\times$ \robot{}}
     &
    \multirow{2.3}{*}{\RoleS{} $\times$ \AerialTechnical{}}
     & LR
     & \textbf{9.316**}
     & -
     & -
    \\
    \arrayrulecolor{gray}
    \cmidrule{3-6}
    \arrayrulecolor{black}
     &
     & coef
     & \textbf{2.866**}
     & -
     & -
    \\ \bottomrule
  \end{tabular}
\end{table*}

\paragraph{\eyeGaze{}}
The proportion of observed social cues that include the eye gaze of each \robot{} is $96.15\%$ for \AerialTechnical{}, $98.08\%$ for \Anthropomorphic{}, $96.79\%$ for \GroundedTechnical{}, and $87.82\%$ for \Zoomorphic{}.
Among the four \robot{}s, the LR test for the mixed models shows a significant effect on the use of eye gaze ($LR=18.212, p < 0.001$).
Participants are less likely to glance at the robot waiter when it is \Zoomorphic{} ($-1.808, p < 0.001$), possibly due to the height of this robot not being in the normal height of human sight.

\paragraph{\upperBody{}}\label{sec:result-upper-body}
Participants rarely move their bodies during the coffee chat. Among all observed social cues, only $8.33\%$ movement of upper bodies for \AerialTechnical{}, $6.41\%$ for \Anthropomorphic{}, $4.49\%$ for \GroundedTechnical{}, and $23.08\%$ for \Zoomorphic{}.
The effects of \Zoomorphic{} on the use of upper body movement are significant ($LR=34.596, p < 0.001$).
Participants move their upper bodies more when interacting with the \Zoomorphic{} robot waiter ($2.170, p < 0.001$), possibly for bending over to interact with it ($19.23\%$).

\subsubsection{\handAndArmGesture{} Feature}
Among all coded \handAndArmGesture{} features, the \role{} has significant effects on \CODEMR{} and \numberOfHands{}, while \AerialTechnical{} and \Zoomorphic{} both have a significant effect on \HH{}.
None of the \role{} or four types of \robot{} has significant effects on other \handAndArmGesture{} features. The results are summarized in Table~\ref{tab:regression-result-gesture}.

\paragraph{\handAndArmGesture{} - \CODEMR{}}
\label{sec:result-gesture-repetition}
The proportion of \handAndArmGesture{} with repetition in the \RoleL{} session is $46.79\%$, and the proportion of \handAndArmGesture{} with repetition in the \RoleS{} session is $51.28\%$. The proportion of \handAndArmGesture{} without repetition in the \RoleL{} session is $39.42\%$, and the proportion of \handAndArmGesture{} without repetition in the \RoleS{} session is $33.01\%$.
The \role{} has a significant effect on the use of \handAndArmGesture{} with repetition ($LR=4.829, p < 0.05$).
When the participants are in the \RoleS{} session, they are less likely to use \handAndArmGesture{} with repetition ($-0.391, p < 0.01$), while they are more likely to use repetitive \handAndArmGesture{} as \RoleL{}.

\paragraph{\handAndArmGesture{} - \HH{}}
The distribution of different \HH{} for different \robot{} is as follows: 
\AerialTechnical{} - $41.67\%$ for between table and head, $32.05\%$ for head level, and $13.46\%$ for above head.
\Anthropomorphic{} - $3.21\%$ for below table, $64.74\%$ for between table and head, $15.38\%$ for head level, and $2.56\%$ for above head.
\GroundedTechnical{} - $0.64\%$ for below table, $73.72\%$ for between table and head, $9.62\%$ for head level, and $0.64\%$ for above head.
\Zoomorphic{} - $30.77\%$ for below table, $50.64\%$ for between table and head, $1.92\%$ for head level, and $0.64\%$ for above head.
The LR test shows a significant effect on \AerialTechnical{} ($LR=44.537, p<0.001$) and \Zoomorphic{} ($LR=36.178, p<0.001$).
The heights of participants' hands are lower when interacting with \Zoomorphic{} ($-1.449, p < 0.001$) robots and higher when interacting with \AerialTechnical{} robots ($LR=1.644, p<0.001$), which is consistent with the heights of these robots.

\paragraph{\handAndArmGesture{} - \numberOfHands{}}
\label{sec:result-gesture-handedness}
The distribution of \handAndArmGesture{} with different \numberOfHands{} used in the \RoleL{} session is $75\%$ for \textit{single-handed} gestures and $11.54\%$ for \textit{two-handed} gestures. 
The distribution of \handAndArmGesture{} with different \numberOfHands{} used in the \RoleS{} session is $77.24\%$ for \textit{single-handed} gestures and $7.05\%$ for \textit{two-handed} gestures. 
The \role{} has a significant effect on the \numberOfHands{} ($LR=5.478, p<0.05$). 
Participants are less likely to use gestures of two hands in the \RoleS{} session ($-0.539, p < 0.05$).

\begin{table*}[htbp]
  \centering
  \caption{Regression coefficients for predicting \handAndArmGesture{} features using Cumulative Link Mixed Models with forward progressive selection using likelihood-ratio Chi-squared tests. For each factor, the baseline is indicated in parentheses. In the table, $***$: $p<0.001$; $**$: $p<0.01$; $*$: $p<0.05$. ``LR'' stands for likelihood-ratio for model selection. Non-significant interaction terms are not shown.}
  \label{tab:regression-result-gesture}
  \begin{tabular}{ccclll}
    \toprule
    \multicolumn{3}{c}{{\handAndArmGesture{} Feature}}
     &
    \repetition
     &
    \handHeight
     &
    \begin{tabular}[l]{@{}l@{}}
      \textit{number-}
      \\
      \textit{of-hands}
    \end{tabular}
    \\
    \midrule
    \multirow{2.3}{*}{\role}
     &
    \multirow{2.3}{*}{
      \begin{tabular}[c]{c}\RoleS{}
        \\ (\RoleL{})
      \end{tabular}}
     &
    LR
     &
    \textbf{4.829*}
     &
    -
     &
    \textbf{5.478*}
    \\
    \arrayrulecolor{gray}
    \cmidrule{3-6}
    \arrayrulecolor{black}
     &
     & coef
     &
    \textbf{-0.391**}
     &
    -
     &
    \textbf{-0.539*}
    \\
    \cmidrule{1-6}
    \multirow{10.5}{*}{\robot}
     &
    \multirow{2.3}{*}{
      \begin{tabular}[c]{c}
        \AerialTechnical
        \\
        (non-\AerialTechnical)
      \end{tabular}}
     &
    LR
     &
    0.530
     &
    \textbf{44.537***}
     &
    0.012
    \\
    \arrayrulecolor{gray}
    \cmidrule{3-6}
    \arrayrulecolor{black}
     &
     & coef
     &
    -
     &
    \textbf{1.644***}
     &
    -
    \\
    \cmidrule{2-6}
     &
    \multirow{2.3}{*}{
      \begin{tabular}[c]{c}
        \Anthropomorphic
        \\
        (non-\Anthropomorphic)
      \end{tabular}}
     &
    LR
     &
    0.214
     &
    0.876
     &
    0.725
    \\
    \arrayrulecolor{gray}
    \cmidrule{3-6}
    \arrayrulecolor{black}
     &
     & coef
     & -
     & -
     & -
    \\
    \cmidrule{2-6}
     &
    \multirow{2.3}{*}{
      \begin{tabular}[c]{c}
        \GroundedTechnical
        \\
        (non-\GroundedTechnical)
      \end{tabular}}
     &
    LR
     &
    1.832
     &
    0.876
     &
    0.113
    \\
    \arrayrulecolor{gray}
    \cmidrule{3-6}
    \arrayrulecolor{black}
     &
     & coef
     & -
     & -
     & -
    \\
    \cmidrule{2-6}
     &
    \multirow{2.3}{*}{
      \begin{tabular}[c]{c}
        \Zoomorphic
        \\
        (non-\Zoomorphic)
      \end{tabular}}
     &
    LR
     &
    2.509
     &
    \textbf{36.178***}
     &
    1.719
    \\ \arrayrulecolor{gray}
    \cmidrule{3-6}
    \arrayrulecolor{black}
     &
     & coef
     & -
     & \textbf{-1.449***}
     & -
    \\\bottomrule
  \end{tabular}
\end{table*}

\subsubsection{\verbal{} Feature}
Among all coded \verbal{} features, \role{} has significant effects on \VPronoun{}, \politeness{} - \VPIV{} and \politeness{} - \VPVIII{}. For \VV{}, the effects of \role{} and \Zoomorphic{} are significant.
A summary of the effects of IVs on verbal features is presented in Table~\ref{tab:regression-result-verbal}.

\paragraph{\verbal{} - \VPronoun{}}
An \VPronoun{} is a deictic expression (\eg this/that/here/there) whose real-world meaning depends on the context.
Among all the social cues, the proportion of \VPronoun{} in the \RoleL{} session is $7.37\%$, and the proportion of \VPronoun{} in the \RoleS{} session is $12.50\%$.
The LR test for the mixed models shows a significant effect of \role{} on the use of pronouns ($LR=7.612, p < 0.01$).
When the participants act as \RoleS{}, they are more likely to reference something unclearly to the robot in their social cues ($0.985, p < 0.01$).

\paragraph{\verbal{} - \politeness{} words}
\label{sec:result-verbal-politeness}
For politeness word \VPIV{}, the result of the LR test shows a significant effect on \role{} ($LR=8.302, p<0.01$) and \GroundedTechnical{} ($LR=3.985, p<0.05$).
$2.88\%$ of the social cues includes \VPIV{} words in their speech content when the participants are the \RoleL{}, and $7.05\%$ of the social cues includes \VPIV{} words when the participants are the \RoleS{}.
The \RoleS{} role uses significantly more \VPIV{} words ($1.487, p < 0.01$).
\VPVIII{} is another way of showing politeness, 
and the \role{} also has a significant effect on the use of \VPVIII{} words ($LR=7.661, p<0.01$).
$2.88\%$ of the social cues includes \VPVIII{} words in their speech content when the participants are the \RoleL{}, and $6.41\%$ of the social cues includes \VPVIII{} words when the participants are the \RoleS{}.
As \RoleS{}, participants are significantly more likely to use \VPVIII{} words ($1.508, p < 0.05$) in their social cues.

\paragraph{\verbal{} - \VV{}}
Both \role{} ($LR=52.485, p < 0.001$) and \Zoomorphic{} ($LR=8.294, p < 0.01$) have significant effects on the verbal \VV{} of the social cues.
The coefficient of \RoleS{} ($1.622, p < 0.001$) shows that participants are more likely to use higher volume in the \RoleS{} session. 
The coefficient of \Zoomorphic{} ($-0.855, p<0.01$) shows that participants are more likely to use lower volume when talking to the \Zoomorphic{} robot.

\begin{table*}[htbp]
  \centering
  \caption{Regression coefficients for predicting \verbal{} features using Cumulative Link Mixed Models with forward progressive selection using likelihood-ratio Chi-squared tests. For each factor, the baseline is indicated in parentheses. In the table, $***$: $p<0.001$; $**$: $p<0.01$; $*$: $p<0.05$. ``LR'' stands for likelihood-ratio for model selection. Non-significant interaction terms are not shown.}
  \label{tab:regression-result-verbal}
  \begin{tabular}{cccllll}
    \toprule
    \multicolumn{3}{c}{{\verbal{} Feature}}
     &
    \begin{tabular}[l]{@{}l@{}}
      unclear
      \\
      reference
    \end{tabular}
     &
    \begin{tabular}[l]{@{}l@{}}
      \politeness{} -
      \\
      \VPIV
    \end{tabular}
     &
    \begin{tabular}[l]{@{}l@{}}
      \politeness{} -
      \\
      \VPVIII
    \end{tabular}
     &
    \VV
    \\
    \midrule
    \multirow{2.3}{*}{\role}
     &
    \multirow{2.3}{*}{
      \begin{tabular}[c]{c}\RoleS{}
        \\ (\RoleL{})
      \end{tabular}}
     &
    LR
     &
    \textbf{7.612**}
     &
    \textbf{8.302**}
     &
    \textbf{7.661**}
     &
    \textbf{52.485***}
    \\
    \arrayrulecolor{gray}
    \cmidrule{3-7}
    \arrayrulecolor{black}
     &
     & coef
     &
    \textbf{0.985**}
     &
    \textbf{1.487**}
     &
    \textbf{1.508*}
     &
    \textbf{1.622***}
    \\
    \cmidrule{1-7}
    \multirow{10.5}{*}{\robot}
     &
    \multirow{2.3}{*}{
      \begin{tabular}[c]{c}
        \AerialTechnical
        \\
        (non-\AerialTechnical)
      \end{tabular}}
     &
    LR
     &
    0.040
     &
    0.631
     &
    1.394
     &
    1.193
    \\
    \arrayrulecolor{gray}
    \cmidrule{3-7}
    \arrayrulecolor{black}
     &
     & coef
     &
    -
     &
    -
     &
    -
     &
    -
    \\
    \cmidrule{2-7}
     &
    \multirow{2.3}{*}{
      \begin{tabular}[c]{c}
        \Anthropomorphic
        \\
        (non-\Anthropomorphic)
      \end{tabular}}
     &
    LR
     &
    1.428
     &
    0.351
     &
    2.924
     &
    <0.001
    \\
    \arrayrulecolor{gray}
    \cmidrule{3-7}
    \arrayrulecolor{black}
     &
     & coef
     & -
     & -
     & -
     & -
    \\
    \cmidrule{2-7}
     &
    \multirow{2.3}{*}{
      \begin{tabular}[c]{c}
        \GroundedTechnical
        \\
        (non-\GroundedTechnical)
      \end{tabular}}
     &
    LR
     &
    0.279
     &
    \textbf{3.985*}
     &
    0.936
     &
    1.163
    \\
    \arrayrulecolor{gray}
    \cmidrule{3-7}
    \arrayrulecolor{black}
     &
     & coef
     & -
     & \textbf{-1.659}
     & -
     & -
    \\
    \cmidrule{2-7}
     &
    \multirow{2.3}{*}{
      \begin{tabular}[c]{c}
        \Zoomorphic
        \\
        (non-\Zoomorphic)
      \end{tabular}}
     &
    LR
     &
    3.009
     &
    0.045
     &
    0.091
     &
    \textbf{8.294**}
    \\ \arrayrulecolor{gray}
    \cmidrule{3-7}
    \arrayrulecolor{black}
     &
     & coef
     & -
     & -
     & -
     & \textbf{-0.855**}
    \\\bottomrule
  \end{tabular}
\end{table*}

\subsection{Interview}

\subsubsection{Considerations on Modality and Articulation}
\label{sec:interview-modality}
Generally, participants would choose minimum articulations for their social cues to guarantee efficiency and minimize cognitive load according to the interview results.
As most explicit social cues include \handAndArmGesture{} or \verbal{} modalities, we would mainly focus on the rationales for choosing these two modalities in the following sections.

\paragraph{\handAndArmGesture{}} 
\label{sec:interview-modality-hand-gesture}
For explicit cues mentioned in the interview, participants chose to use the \handAndArmGesture{} modality the most when signaling the robot waiter during the conversation (Section~\ref{sec:result-overall}). 
The rationale is that they are more familiar with the \handAndArmGesture{}, and it is natural and intuitive for them to use common gestures (U04, U09, U15, U22). 
Many participants reckoned that the robot waiter should recognize commonsense gestures like \dismissiveWave{} for ``Go away'', \beckoningWave{} for ``Come'', and \textit{thumbs up or down} for good or bad feedback (U07, U21, U22).
For more complex situations, some of them assumed that the robot waiter should understand the context and the intention behind the gestures (U04, U05, U22), such as whether the \textit{pointing to a drink} gestures mean that the robot should offer the drink or collect the empty cup (U13, U19, U21, U23, U24), or the \wave{} gesture to express ``No'' or ``Wrong''.

\paragraph{\verbal{}} 
\label{sec:interview-modality-verbal}
Participants mentioned that they would use verbal cues when they want to express their intentions clearly and efficiently (U04, U07, U12).
Some of them mentioned that they think gestures cannot properly express some of their intentions, so they would use verbal cues to clarify (U15, U19, U22); for example, they would \ReferenttwoI{} with the \wave{} gesture and the verbal cue ``Fanta'' to the robot waiter with the wrong order.
Meanwhile, U20 mentioned that they were not used to speaking to the robot in public; thus, they did not use verbal cues during the experiment.

\paragraph{Other Modalities}
\label{sec:interview-modality-other}
We also asked participants about their eye gaze if we noticed that their gaze turned to the referent-related objects or the robot waiter during the conversation. 
Most participants said their gaze unconsciously followed their social cues (U01, U11, U22). 
Some mentioned that they were attracted by the robot waiter's movement even if they did not intend to signal it (U06, U09, U12). 
Some said they kept looking at the robot waiter to make sure it understood their signals and did not do anything unexpected (U05, U11, U18).
The \headMotion{} modality mainly goes with gaze, or some participants would nod to \ReferentoneI{} or \ReferentthreeIV{} to the robot (U01, U03, U08, U16, U20).
As for other modalities, participants mentioned that they would prefer not to use \facialExpression{} to signal the robot waiter since they thought it was not socially appropriate (U06, U11, U15) or they felt it was difficult to detect (U17, U21), and only a few participants mentioned smiling, raising their eyebrows, or blinking to signal awareness or confirmations to the robot waiter (U17, U20).
And they would not use \legAndFoot{} to signal the robot waiter since they were sitting and did not have the habit of using their legs or feet to signal others (U20, U22), and U21 felt it was not socially appropriate to use \legAndFoot{}. 
The only participant (U15) using \legAndFoot{} mentioned that they had an experience participating in a project that collected their foot signals, and it reminded them to use their legs to signal the robot waiter.

\subsubsection{Conversation Role}
\label{sec:interview-role}
Participants generally felt more cognitively demanding when they were the \RoleS{} in the conversation, so they would choose the most simple social cue or even ignore the robot to avoid distractions when they were in the middle of presenting something to the potential employers ({U19, U24}).
Some participants mentioned that they would use more verbal cues when they were the \RoleS{} in the chat interrupted by the robot waiter ({U03, U05, U07, U08, U10}). 
Since they managed the conversation flow, when they could pause the topic, they quickly clarified their intentions to the robot waiter with simple commands.
Compared to gesturing, some of them felt speaking was a more straightforward way to instruct the robot and minimize distractions since they were speaking and did not worry about interrupting the potential employers in the conversation ({U07, U08}).
As for the \RoleL{} session, many participants mentioned that they would avoid speaking to the robot waiter to avoid interrupting the conversation, and they would use more complex gestures to signal the robot waiter since they felt that they could leave the conversation for a little while to settle down the robot interruptions when they were not talking in the coffee chat ({U07, U08, U19, U21}).
Nevertheless, a few participants mentioned that they would still use verbal cues when they were the listener in the conversation ({U02, U06}). 
They would turn down their volume to quickly instruct the robot waiter in the \RoleL{} session, while they said that they did not have the cognitive load to use verbal cues to signal the robot in the \RoleS{} session.
Overall, some participants mentioned that they thought it was more appropriate for the robot to find the right people who were not speaking to interact with if the waiter had to interrupt the conversation ({U09, U14, U22}).

\subsubsection{Robot Perception}
\label{sec:interview-robot-perception}
The \AerialTechnical{} and \GroundedTechnical{} robot waiters were perceived as machines or tools (U08, U12), \Zoomorphic{} was perceived as more playful to attract customers (U12, U17), and \Anthropomorphic{} was perceived more like a human waiter (U07, U20).
U18 said they would be more strict with the \Anthropomorphic{} robot since they thought if the robot looked like humans and did not perform the tasks as well as humans, then it would disappoint them. 
Given their different perceptions, most participants did not think that the appearance of the robot waiter would affect their choice of social cues, but only the physical properties such as heights or, more specifically, the sensor positions would affect their social cues (U07, U08, U10, U19, U21, U22, U24).
For example, they perceived that the cameras of the robot waiters were in the front or on the ``face'', so they would gesture at a higher position to \AerialTechnical{} and \Anthropomorphic{} robots, and blend over to gesture at a lower position to the \Zoomorphic{} robot (U21, U22, U24).
Similarly, some of them felt that the robot waiter could not ``hear'' them clearly if their perceived microphones of the robot were too far away, so they either chose not to use verbal cues unless the robot was closer (U11) or leaned their bodies to get closer to the robot waiter to speak to it (U13, U22, U23).
Moreover, the participants' perceptions of the robots' capability and intelligence would also affect their behaviors during the interaction.
Many participants complained about the interruptions caused by the robot waiter, which reduced their willingness to interact with the robot (U01, U11).
Some thought the robot waiter was not responding to their social cues, so they may repeat their cues several times or do the job themselves instead of signaling the robot to do it (U13, U14).
Some participants mentioned that they felt the robot waiter was not intelligent enough to understand their social cues, so they would use more explicit cues to signal the robot waiter (U02, U13).
Given an emergency error on purpose in our referents (\ReferenttwoII), only one participant (U11) mentioned feelings of unsafe when interacting with the \AerialTechnical{} robot during the experiment due to the experience with an unsteady drone.
That participant kept a safe distance while interacting with the \AerialTechnical{} robot.
This suggests that although some participants mentioned that they might change their behavior if the robot waiters become more dangerous (U11, U20), most participants may feel safe when interacting with them during the experiment.

\subsubsection{Social Context}
\label{sec:interview-social-context}
Most participants would consider the politeness and the social appropriateness of their social cues. 
For the \handAndArmGesture{} modality, they would not use the \dismissiveWave{} or \textit{finger pointing} to the direction of the two potential employers to avoid misunderstanding (U19, U20, U22).
Some of them would hide their gestures below the table or near their body to signal the robot inconspicuously (U01, U12, U19).
For the \verbal{} modality, some participants would cover their {mouth} and whisper to the robot to avoid interrupting the speakers (U05, U19, U22), and one (U16) hoped that the robot waiter could read their lip so that they could signal the robot without speaking out loud.
Some of them said that they would use polite words to instruct the robot waiter either because they politely treat the robot waiter the same as human waiters (U05, U11, U12, U14, U21), or because they want to show their etiquette in front of two potential employers (U12, U20), while some took the robot waiter as a machine or a tool and thus use simple and direct words to instruct the robot waiter (U02, U06, U15, U19, U24).
Despite the simulated social scenario in our experiment, participants generally reflected that the chosen public space was appropriate for a coffee chat (U04, U05, U07, U08, U09, U14, U15, U17, U21, U22, U23), while some (U08, U13, U18) mentioned that the environment was quieter than coffee chats they experienced and some (U11, U12) said they would adjust the volume and subtlety of their social cues according to the background music or noise level of the environment.

\subsubsection{Other Rationale and Comment}
\label{sec:interview-other}
Participants mentioned personal reasons for choosing some social cues, 
for example, some were more introverted, so they would choose inconspicuous cues to signal the robot waiter and avoid speaking (U07, U12, U20), 
and some may prefer ignoring the robot waiter in most cases to avoid interruptions and distractions (U02, U03).
Some participants kept using similar gestures for different referents based on their habits; for example, U08 waved his hand in all dimensions throughout both sessions, and U22 used the same \GNIV{} gesture for a large portion of referents. 
Several participants said they preferred to rate on a touch screen rather than use social cues to signal the robot waiter (U02, U06, U19, U23), and some hoped there were buttons on the table so that they could press to call over the robot waiter (U16, U17) or specify the service type (U10).

\section{Discussion}

\subsection{Considerations on Human Social Cues for HRI in Social Contexts}

Our work provides empirical insights showing how human perceptions of robots, the primary task, and social context affect human social cues in human-robot side interactions. 
We {discuss} the implications and generalizations of our findings from the perspectives of humans, robots, and the social context.

\subsubsection{Choice of Social Cues and Human Intention}

Social cues are usually the most efficient way of communication when interacting with a service robot as a side task, according to the qualitative results in Section~\ref{sec:interview-modality}.
Our participants' top explicit choices of \handAndArmGesture{} and \verbal{} modalities are consistent with the common channels of human conversational interaction \cite{quekMultimodalHumanDiscourse2002}.
We identified some frequently used articulations of explicit cues, such as pointing, waving, and imperative sentences.
We found people may use similar gestures for different intentions and expect the robot to understand the context to disambiguate the intentions. At the same time, there were also variances in their choice of social cues according to personal habits and preferences.
These findings suggest that the design of social service robots should consider the common patterns of human social cues to understand user intentions, as well as individual differences in human social behaviors to provide personalized services.

\subsubsection{Perception on Robot}

Interview analysis (\Cref{sec:interview-robot-perception}) indicates that robot morphology does not directly determine social cue selection but shapes interaction behaviors through user perceptions. 
The reported influences of perceptions mainly lie in the following dimensions: perceived physical capability, perceived safety, and perceived intelligence,
which are consistent with \cite{bartneckMeasurementInstrumentsAnthropomorphism2009b} that robot morphology and behavior are two main factors that affect human perceptions (i.e., anthropomorphism, animacy, likability, perceived intelligence, and perceived safety) of robots.

\textit{Perceived Physical Capability.}
The interview results show that participants' assumptions about the robots' physical embodiments and capabilities may affect their behaviors and social cues.
Their assumptions usually follow common \textit{anthropomorphic} patterns. 
Participants inferred sensor placement (e.g., cameras as ``heads'' for \Anthropomorphic{} and \GroundedTechnical{} robots vs. front-facing for \Zoomorphic{} and \AerialTechnical{} designs) based on perceived embodiments and capabilities, influencing social cue adjustments. 
These findings align with \cite{onnaschTaxonomyStructureAnalyze2021} that the morphology shapes a user's expectations of the robot's functioning, 
and they can be further connected to the choice of social cues, such that participants adjust their social cues based on the perceived robot's physical capability to make the interactions effective.
Our quantitative results verify that the robot morphology has significant effects on \HH{}, \VV{}, and \textit{upper body motion} (\Cref{sec:result-effects}).
These findings highlight the importance of sensor design and the alignment of robot capabilities with users' assumptions.
To collect valid and high-quality human social cues for better recognition and understanding, sensor placements should align with user expectations or be explicitly communicated to ensure effective interaction. 

\textit{Perceived Safety.}
According to the participants' familiarities with the four morphologies (\Cref{method:elicitation-study:participants}) and the interview (\Cref{sec:interview-robot-perception}), most participants have heard of the four morphologies of social service robots, and many of them have interacted with them in canteens, hotels, or other places for entertainment before.
Participants' general positive experience with these service robots and the virtual prototyping of the robots in the experiment may contribute to most participants' feeling safe in the experiments (\Cref{sec:interview-robot-perception}).
Nevertheless, safety perceptions was demonstrated through proxemic adjustments: more than $8\%$ of observed cues involved leaning away from the robot during \ReferenttwoII{} interactions (\Cref{tab:top_cues_by_referents}), while \AerialTechnical{} robots elicited increased \upperBody{} cues. 
These findings complement the design of \citet{mandalUsingProxemicsCorrective2024} that robots can utilize the human proxemics as a corrective feedback signal.
To better understand how each robot morphology affects human perceived safety and the choice of social cues, 
future research can consider explicitly asking participants about their perceptions of each robot morphology's safety, given their experience with the robots, and examining the relationship between the perceived safety and the choice of social cues.

\textit{Perceived Intelligence.}
The interview results show that participants' perceptions of the intelligence of the robots may influence human's willingness to interact with the robot, i.e., they may show more impatient behaviors in their explicit or implicit social cues.
From appearance only, the \Anthropomorphic{} robots are expected to be more intelligent than \Zoomorphic{}, \GroundedTechnical{} and \AerialTechnical{} robots according to the interview (\Cref{sec:interview-robot-perception}).
However, if their behaviors are not aligned with the participants' expectations of their intelligence, they may become more impatient when interacting with them.
The variation of perceived intelligence is aligned with the findings from \citet{tusseyevaPerceivedIntelligenceHumanRobot2024}'s survey.
Therefore, it is important to design the robot behaviors to match the participants' expectations of their intelligence to maintain the participants' willingness to interact with the robots.

\subsubsection{Primary Task and Social Context}
\label{sec:discussion-social-context}
The primary task and social context also affect human social cues in human-robot side interactions.
People may adjust their social cues based on the primary task they are engaged in, as well as the social context they are in.
According to our qualitative results (\Cref{sec:interview-social-context}), three main factors play important roles in the choice of social cues: their cognitive load, social attention received, and the social appropriateness of the cues.
The significantly more choice of \verbal{} cues in the \RoleS{} condition (\Cref{sec:result-modality-verbal}), as well as the significantly more complex \handAndArmGesture{} (\CODEMR{} in \Cref{sec:result-gesture-repetition}, and \CODEH{} in \Cref{sec:result-gesture-handedness}) in the \RoleL{} condition, may be due to the higher cognitive load in the \RoleS{} condition according to the interview.
Thus, highly mentally demanding tasks may lead to simpler and more direct social cues to avoid distractions.
At the same time, people usually choose more socially appropriate cues, especially when they receive more social attention, echoing the results of significantly more politeness words chosen in the \RoleS{} session (\Cref{sec:result-verbal-politeness}).
Therefore, social service robots need to understand the interaction context, including the surrounding environment, the primary task, and the social context, to provide appropriate services and receive detailed instructions and feedback from the users.
In situations that require immediate human responses, social service robots need to be equipped with the ability to understand users' intentions, given the minimum set of social cues.

\subsection{Design Implications for Social Service Robot Interactions}
\subsubsection{Robot Approaching Strategy}

People's general unwillingness to interact with the robot waiter in the \RoleS{} session and complaints about the interruptions of the robot waiter suggest that the approaching strategy is important for the robot to initiate interactions with users (\Cref{sec:interview-robot-perception}).
The robot should find the right time to serve the users, especially when they are engaged in their main tasks.
In terms of the conversation in our work, the pauses between conversation turns may be good timing for the robot to approach the users according to the previous HRI research \cite{nagao1994SocialInteractionMultimodal,palinkoHowShouldRobot2018}.
Nevertheless, in cases that require immediate human feedback, finding the right person to interact with when serving a group of people is also important to minimize the interruption.
The cognitive load of the users discussed in \Cref{sec:discussion-social-context} suggests that it is better for the robot to approach the person who is less involved in the primary task, where detecting human's cognitive load is a potential challenge and future design considerations.

\subsubsection{Human Social Cue Processing}

The general patterns of explicit and implicit human social cues (\Cref{sec:result-overall}) suggest that the robot should be able to process all modalities of cues to better understand human intentions.
The robot should be able to deduce the human willingness to interact, the instructions they signaled, and the feedback they provided.
The main challenge would be learning the commonsense semantics of human social cues and distinguishing similar social cues in different contexts (\Cref{sec:interview-modality}).
One potential consideration based on our findings is that the robot should be able to learn from implicit cues, such as gaze direction or implicit facial expressions, to disambiguate human intentions, emphasize the important parts, and determine the scope of human feedback.
These implicit cues for robots are functioning as expressing emotion and sending relational messages according to \cite{vinciarelliSocialSignalsTheir2008}, which are important parts of human social communications and necessary to be considered in the robot design in human social cue processing.

\subsubsection{Robot Response}

When it receives human social cues and understands human intentions, the robot should be able to respond appropriately to human needs.
According to the interview, a slow response will increase the participants' impatience and decrease their willingness to interact with the robot (\Cref{sec:interview-robot-perception}), which aligns with \cite{kangRobotFeedbackDesign2024}.
The robot should be able to respond in a timely and socially appropriate manner to maintain human engagement and interaction. 
The response should be intuitive and easy to understand while guaranteeing minimum interruptions to humans.
Moreover, the robot should be able to respond to human feedback and instructions and take appropriate actions to fulfill human needs, which highly depends on the recognition and disambiguation of human social cues.
Strategies to repair potential failures are also important considerations.
On the one hand, when intention recognition fails, the service robot may need to initiate additional interactions with humans to confirm.
On the other hand, the interruptions caused by the social service robot service can annoy the users.
How the service robot repairs the relationship with the users to regain their trust \cite{esterwoodLiteratureReviewTrust2022} after undesirable situations is also a key consideration for the robot response strategy.

\subsection{Limitation and Future Work}
Our work has several limitations. 
First, we used an augmented reality (AR) headset to simulate the robot waiter in the given social context.  
Previous study \cite{kamideDirectComparisonPsychological2014} suggests the potential of using AR simulations in experiments that require high controllability and comparable personal spaces and human behaviors towards virtual and real robot prototypes.
Thus, using AR to simulate robots or other experiment settings is a good choice for prototype experiment settings to provide insights into understanding users' behavior and implicating the designs. Still, real tests on robots or products are needed to ground the results in real-world applications.
Second, our study had a relatively small sample size.  
As a result, the independent and identically distributed sample size for each unique condition is small, making it insufficient for a statistical test of the independence between conditions in the same independent variable. 
Instead, we relied on the mixed effect analysis of each variable (given a large enough overall sample size with repeated measures) and qualitative analysis of the interview to understand how different factors may affect human decisions and rationales for choosing social cues to signal the robot in the scope of this paper. 
Third, the participants were mainly students from local universities, which may limit the generalizability of the results to other populations, such as older adults who are less active in different body parts \cite{biswasAreOlderPeople2020, morillo-mendezAgeRelatedDifferencesPerception2024} or people from a different culture.
Fourth, due to the complexity of the social context, we focused only on one scenario of social interaction (i.e., coffee chat), which may not cover all referents and the corresponding social cues in other scenarios involving service robots. 
In addition, according to the suggestions of our pilot study participants, we chose the sitting pose for their comfort, which may limit the use of cues involving lower limb motion, whole body movement, or change in proximity. 

Future work may recruit a larger number of participants from more diverse backgrounds to explore the influence of various personal and cultural factors in human-robot communication.
Further exploration may consider adding standing pose, other social arrangements, and richer service scenarios to acquire a more comprehensive understanding of human interactions with robots in the social context.
Processing and learning from the collected social cue data is another potential direction for future work, aiming to equip service robots with the ability to understand and tackle users' preferences and needs in real contexts.

\section{Conclusion}

This work aims to gain empirical insights into how robot morphology and human leading task roles impact humans' choice of social cues to express their intentions in social encounters.
We conducted an elicitation study with $24$ participants in a simulated coffee chat scenario with potential employers, where participants interacted with four different robots (\AerialTechnical{}, \Anthropomorphic{}, \GroundedTechnical{}, and \Zoomorphic{}) in two different roles (\RoleS{} and \RoleL{}).
Our study collected $624$ for observed social cues, including detailed coded articulations and features for each social cue.
The statistics and quantitative results reveal patterns in human social cues' modalities, articulations, and features. 
Additionally, qualitative analysis of the interview supports the quantitative results. It provides a deep insight into the participants' rationale in choosing the social cues and comments on the robots' appearance, behavior, and social context.
The conversational role significantly affects the adoption of verbal cues, verbal features, and gesture features, mainly because the two roles impose different cognitive loads and social norms. 
The robot morphology significantly affects the adoption of different modalities, gesture features, and verbal features, mainly due to the robot's appearance (\,  e.g., height and similarity to humans).
From these findings, we identify implications for understanding human social cues and inform robot design. 
For example, service robots with distinct morphologies should account for intuitive adjustments users make based on robot appearance, such as designing sensor placements that align with users' expectations. 
Additionally, service robots operating in professional or social contexts should adopt response strategies that minimize disruption, such as detecting subtle gazes or hand gestures and responding with polite, context-sensitive feedback.
These design considerations provide actionable insights for service robots, particularly in their approach strategies, interpretation of human social cues, and interaction responses, to better integrate into dynamic human-centric environments.

\begin{acks}
    We thank our participants for their time, effort, and valuable input; and our actors for their excellent performance in the elicitation study.
    We also thank the anonymous reviewers for their feedback.
    This work is supported by the Research Grants Council of the Hong Kong Special Administrative Region under General Research Fund (GRF) with Grant No. 16207923.
\end{acks}

%%
%% The next two lines define the bibliography style to be used, and
%% the bibliography file.
\bibliographystyle{ACM-Reference-Format}
\bibliography{reference}

\newpage
%%
%% If your work has an appendix, this is the place to put it.
\appendix

\end{document}